\documentclass[preprint,superscriptaddress,nofootinbib]{revtex4}
\usepackage{amssymb}
\usepackage{amsmath}
\usepackage{epsfig}
\usepackage{graphicx}
\usepackage{here}
\usepackage{color}
\def\lapproxeq{\lower .7ex\hbox{$\;\stackrel{\textstyle <}{\sim}\;$}}
\def\gapproxeq{\lower .7ex\hbox{$\;\stackrel{\textstyle >}{\sim}\;$}}

\begin{document} 
%\begin{titlepage}
\begin{flushright}
%September, 2005
\today
\end{flushright}
%\maketitle
%
%\rightline{TODAY}
%\twocolumn
%
%\vspace*{0.2cm}
%
%\begin{center} \begin{large}
\title{
$\tau$- and  $\mu$-physics in a general two Higgs doublet model \\ with $\mu-\tau$ flavor violation}
%}
%\end{large} \end{center}

%\begin{center}
%\vspace*{0.2cm} 
\author{Yuji Omura}
\affiliation{Kobayashi-Maskawa Institute for the Origin of Particles and the Universe,
  Nagoya University, Nagoya, 464-8602, Japan}
\author{Eibun Senaha}
\affiliation{Center for Mathematics and Theoretical Physics and Department of Physics,
  National Central University, Taoyuan, Taiwan 32001, R.O.C.}
\author{Kazuhiro Tobe}
\affiliation{Department of Physics,
          Nagoya University,
      Nagoya, 464-8602, Japan}
\affiliation{Kobayashi-Maskawa Institute for the Origin of Particles and the Universe,
          Nagoya University,
      Nagoya, 464-8602, Japan}
%\vspace*{0.2cm}
%}
%\end{center}
%
%\begin{center}
%
%\affiliation{
%Department of Physics, Nagoya University, Nagoya, Japan
%%
%}
%\end{center}
%
\begin{abstract}
  Motivated by the recent CMS excess in a flavor violating Higgs decay $h\rightarrow \mu\tau$
  as well as the anomaly of muon anomalous magnetic moment (muon g-2), we consider a scenario
  where both the excess in $h \rightarrow \mu \tau$ and the anomaly of muon g-2 are explained
  by the $\mu-\tau$ flavor violation in a general two Higgs doublet model.
  We study various processes involving $\mu$ and $\tau$, and then discuss the typical predictions and
  constraints in this scenario. Especially, we find that the prediction of $\tau \rightarrow \mu \gamma$ can be
  within the reach of the Belle II experiment.
  We also show that the lepton non-universality between $\tau \rightarrow \mu \nu \bar{\nu}$ and
  $\tau \rightarrow e \nu \bar{\nu}$ can be sizable, and hence the analysis of the
  current Belle data and the future experimental improvement would have an impact on this model.
  Besides, processes such as
  $\tau \rightarrow \mu l^+ l^-~(l=e,~\mu)$, $\tau \rightarrow \mu\eta$, $\mu \rightarrow e\gamma$,
  $\mu \rightarrow 3e$, and muon EDM can be accessible, depending on the unknown Yukawa couplings.
  On the other hand, the processes like $\tau \rightarrow e\gamma$ and $\tau \rightarrow e l^+ l^-~  (l=e,~\mu)$
  could not be sizable to observe because of the current strong constraints on the $e-\mu$ and $e-\tau$ flavor violations.
  Then we also conclude that contrary to $h \rightarrow \mu \tau$ decay mode, the lepton flavor violating
  Higgs boson decay modes $h\rightarrow e\mu$ and $h\rightarrow e\tau$ are strongly suppressed, and
  hence it will be difficult to observe these modes at the LHC experiment.
  
\end{abstract}
%\end{titlepage}
%\begin{document}
%
\maketitle
%\flushbottom
%
%
%\vspace*{1cm}
%

\section{Introduction}
The Standard Model (SM) of the elementary particles describes particle physics phenomena remarkably well
up to the electroweak scale. In addition, the recent discovery of a Higgs boson at the
LHC~\cite{Aad:2012tfa,Chatrchyan:2012xdj} strengthens the success of the
SM. On the other hand, the detailed measurements of the Higgs boson properties have just started,
and the whole structure of the Higgs sector may have not been unveiled. Therefore, theoretical
and experimental studies of the extended Higgs sector would be important to understand the nature of
the Higgs sector.

One of simple extensions of the Higgs sector in the SM is a two Higgs doublet model (2HDM) where additional Higgs doublet is introduced and both Higgs doublets can couple to all fermions. As a result, flavor violating phenomena mediated
by the Higgs bosons are predicted~\cite{Bjorken:1977vt}. In most cases, such a flavor violation
has been considered to be avoided because of lack of the experimental supports for the anomalous flavor-violating
phenomena~\cite{Glashow:1976nt,McWilliams:1980kj,Shanker:1981mj,Cheng:1987rs}.

However, the CMS collaboration has recently reported an event excess in a flavor-violating
Higgs decay $h\rightarrow \mu\tau$~\cite{Khachatryan:2015kon}, and it suggests that the best fit
value of the branching ratio is
\begin{eqnarray}
{\rm BR}(h\rightarrow \mu\tau)=(0.84^{+0.39}_{-0.37})~\%,
\end{eqnarray}
where the final state is a sum of $\mu^+\tau^-$ and $\mu^-\tau^+$, and the deviation from the
SM prediction is $2.4\sigma$. 
In addition, the result of the ATLAS experiment has also appeared recently~\cite{ATLAS_hmt},
and it is shown as
\begin{eqnarray}
{\rm BR}(h\rightarrow \mu\tau)=(0.77 \pm 0.62)~\%,
\end{eqnarray}
which is consistent with the CMS result within $1\sigma$. Although these results have not been
conclusive yet, these become strong motivations to study the flavor violating phenomena
predicted by the Beyond Standard Models
\cite{Sierra:2014nqa,Heeck:2014qea,Crivellin:2015mga,deLima:2015pqa,Omura:2015nja,Das:2015kea,Crivellin:2015lwa,Das:2015zwa,Yue:2015dia,Bhattacherjee:2015sia,Mao:2015hwa,He:2015rqa,Goto:2015iha,Chiang:2015cba,Crivellin:2015hha,Cheung:2015yga,Botella:2015hoa,Liu:2015oaa,Baek:2015mea,Huang:2015vpt,Baek:2015fma,Arganda:2015uca,Aloni:2015wvn}.
\footnote{The lepton flavor violating Higgs decays have been investigated before the CMS excess has been reported\cite{Assamagan:2002kf,Brignole:2003iv,Kanemura:2004cn,Arganda:2004bz,
    Kanemura:2005hr,Blankenburg:2012ex,Harnik:2012pb}.}
%%%%%%%%%%%%%%%%%%%%%%%%%%%%%%%%%
%%%%%%%%%%%REFERENCE%%%%%%%%%%%%%%%%%%%%%

In Ref.~\cite{Omura:2015nja}, we pointed out a possibility that the $\mu-\tau$ flavor violation in
general 2HDM can explain not only the CMS excess in the Higgs decay $h\rightarrow \mu\tau$, but also
the anomaly of muon anomalous magnetic moment (muon g-2)~\cite{Agashe:2014kda}. This possibility is interesting because two
unexplained phenomena can be accommodated in the 2HDM, and hence it is worth further investigating
this possibility.
In this paper, we study phenomena related to $\mu$ and $\tau$ lepton physics in the scenario
to see whether there are any interesting predictions and constraints caused by the $\mu-\tau$ flavor violation.

The paper is organized as follows. In section II, we present a general 2HDM where both Higgs doublets
couple to all fermions. We discuss the Yukawa interactions and Higgs mass spectrum in the model.
In section III, we consider a solution where the CMS excess in $h \rightarrow \mu \tau$ decay as well as
muon g-2 anomaly can be explained by the $\mu-\tau$ flavor violating Yukawa interactions in the model.
We show the typical parameter space where both anomalies are explained.
In section IV, we discuss $\tau$- and $\mu$-physics in the interesting region studied in the previous section.
Especially, we study $\tau \rightarrow \mu \gamma$, $\mu \rightarrow e \gamma$, muon electric dipole moment
(muon EDM), $\tau \rightarrow \mu \nu \bar{\nu}$, $\tau^- \rightarrow \mu^- l^+ l^-$ ($l=e,~\mu$),
$\mu^+\rightarrow e^+ e^- e^+$, and $\tau \rightarrow \mu \eta$.
The prediction of $\tau \rightarrow \mu \gamma$ can be within the reach of the Belle II experiment, which will start near future. 
The extra Higgs boson correction to $\tau \rightarrow \mu \nu \bar{\nu}$ can be as large as $10^{-3}-10^{-4}$, but
it is not large in $\tau \rightarrow e \nu \bar{\nu}$ mode.
The future improvement of measurement on lepton flavor universality in $\tau \rightarrow \mu (e)\nu \bar{\nu}$ decay will be important in the scenario.
We also find that unlike the $\mu-\tau$ flavor violation suggested by the CMS result, the $e-\tau$ and $e-\mu$
flavor violations are severely constrained since the constraint from the $\mu \rightarrow e \gamma$ is strong.
Therefore, the processes related to $e-\tau$ and $e-\mu$ flavor violations are suppressed.
In section V, we also discuss an implication to Higgs physics. Since $e-\tau$ and $e-\mu$ flavor violating
Yukawa couplings are strongly suppressed, $h\rightarrow e\tau$ and $h \rightarrow e\mu$ Higgs decay modes
will not be observed in this scenario, contrary to $h\rightarrow \mu\tau$ mode.
In section VI, we summarize our results.

%%%%%%%%%%%%%%%%%%%%%%%%%%%%%%%%%%%%%
\section{General two Higgs doublet model}
In a general two Higgs doublet model, there are no symmetry to
distinguish the two different Higgs doublets. Thus, both the Higgs doublets can couple
to all fermions, and hence there are flavor violating interactions in Higgs sector.
In general, when the Higgs potential is minimized in the SM-like vacuum,
both neutral components of Higgs doublets develop nonzero vacuum expectation values (vevs). Taking a certain
linear combination, we can define the basis where only one of the Higgs doublets
obtains the nonzero vev as follows:
\begin{eqnarray}
  H_1 =\left(
  \begin{array}{c}
    G^+\\
    \frac{v+\phi_1+iG}{\sqrt{2}}
  \end{array}
  \right),~~~
  H_2=\left(
  \begin{array}{c}
    H^+\\
    \frac{\phi_2+iA}{\sqrt{2}}
  \end{array}
  \right),
\label{HiggsBasis}
\end{eqnarray}
where $G^+$ and $G$ are Nambu-Goldstone bosons, and $H^+$ and $A$ are a charged Higgs boson and a CP-odd
Higgs boson, respectively. Then $H^+$ and $A$ are in the mass eigenstates. 
The CP-even neutral Higgs bosons $\phi_1$ and $\phi_2$ can mix and form mass
eigenstates, $h$ and $H$ ($m_H>m_h$), in general:
\begin{eqnarray}
  \left(
  \begin{array}{c}
    \phi_1\\
    \phi_2
  \end{array}
  \right)=\left(
  \begin{array}{cc}
    \cos\theta_{\beta \alpha} & \sin\theta_{\beta \alpha}\\
    -\sin\theta_{\beta \alpha} & \cos\theta_{\beta \alpha}
  \end{array}
  \right)\left(
  \begin{array}{c}
    H\\
    h
  \end{array}
  \right).
\end{eqnarray}
Here $\theta_{\beta \alpha}$ is the mixing angle and fixed by the Higgs potential analysis. Note that when $\cos\theta_{\beta \alpha}\rightarrow 0$
  ($\sin\theta_{\beta\alpha}\rightarrow 1$), the interactions of $\phi_1$ approach to those of the SM Higgs boson.

\subsection{Yukawa interactions}
In mass eigenbasis for the fermions, the Yukawa interactions are expressed by
\begin{eqnarray}
  {\cal L}&=&-\bar{Q}_L^i H_1 y^i_d d_R^i -\bar{Q}_L^i H_2 \rho^{ij}_d d_R^j
  %\nonumber \\
  %&&
  -\bar{Q}_L^i (V_{\rm CKM}^\dagger)^{ij}\tilde{H}_1 y^j_u u_R^j -\bar{Q}_L^i
  (V_{\rm CKM}^\dagger)^{ij}\tilde{H}_2
  \rho^{jk}_u u_R^k \nonumber\\
  &&-\bar{L}_L^i H_1 y^i_e e_R^i -\bar{L}_L^i H_2 \rho^{ij}_e e_R^j,
\end{eqnarray}
where $i$ and $j$ represent flavor indices, and $Q=(V_{\rm CKM}^\dagger u_L,d_L)^T$,
$L=(V_{\rm MNS} \nu_L, e_L)^T$ are defined. $V_{\rm CKM}$ is the Cabbibo-Kobayashi-Maskawa (CKM) matrix
and $V_{\rm MNS}$ is the Maki-Nakagawa-Sakata (MNS) matrix.
Fermions $(f_L,~f_R)$ $(f=u,~d,~e,\nu)$ are mass eigenstates.
Here we have assumed that the tiny neutrino masses are achieved by the seesaw mechanism introducing
super-heavy right-handed neutrinos, so that in the low-energy effective theory, the left-handed
neutrinos have a $3\times 3$ Majorana mass matrix.
The Yukawa coupling constants $\rho_f^{ij}$ are general $3\times3$ complex matrices and
can be sources of the Higgs-mediated flavor changing processes.

In mass eigenstates of Higgs bosons, the Yukawa interactions are given by
\begin{align}
  {\cal L}&=-\sum_{f=u,d,e}\sum_{\phi=h,H,A} y^f_{\phi i j}\bar{f}_{Li} \phi f_{Rj}+{\rm h.c.}
  \nonumber\\
  &\quad -\bar{\nu}_{Li} (V_{\rm MNS}^\dagger \rho_e)^{ij}  H^+ e_{Rj}
  -\bar{u}_i(V_{\rm CKM}\rho_d P_R-\rho_u^\dagger V_{\rm CKM} P_L)^{ij} H^+d_j+{\rm h.c.},
\end{align}
where
\begin{align}
  y^f_{h\;ij}&=\frac{m_{f}^i}{v}s_{\beta\alpha}\delta_{ij}+\frac{\rho_{f}^{ij}}{\sqrt{2}} c_{\beta\alpha},
  \nonumber\\
  y^f_{H\;ij}&=\frac{m_f^i}{v} c_{\beta \alpha}\delta_{ij}-\frac{\rho_f^{ij}}{\sqrt{2}} s_{\beta\alpha},
  \nonumber \\
  y^f_{A\;ij}&=
  \left\{
  \begin{array}{c}
    -\frac{i\rho_f^{ij}}{\sqrt{2}}~({\rm for}~f=u),\\
    \frac{i\rho_f^{ij}}{\sqrt{2}}~({\rm for}~f=d,~e)
  \end{array}
  \right.
  \label{yukawa}
\end{align}
are defined with $c_{\beta\alpha}\equiv \cos\theta_{\beta\alpha}$ and $s_{\beta\alpha}\equiv\sin\theta_{\beta\alpha}$.
%, which is diagonalized by the $V_{\rm MNS}$
%matrix.
Note that when $c_{\beta\alpha}$ is small, the Yukawa interactions of $h$ are almost equal to those of
the SM Higgs boson, however, there are small flavor violating interactions $\rho_f^{ij}$ which are suppressed
by $c_{\beta\alpha}$. On the other hand, the Yukawa interactions of heavy Higgs bosons ($H$, $A$, and $H^+$)
mainly come from the $\rho_f^{ij}$ couplings.

\subsection{Higgs mass spectrum}
Let us comment on the relation between the Higgs masses and the parameters in the Higgs potential. The renormalizable Higgs potential in the general 2HDM is given by
\begin{align}
  V&=M_{11}^2 H_1^\dagger H_1+M_{22}^2 H_2^\dagger H_2-\left(M_{12}^2H_1^\dagger H_2+{\rm h.c.}
  \right)\nonumber \\
&+\frac{\lambda_1}{2}(H_1^\dagger H_1)^2+\frac{\lambda_2}{2}(H_2^\dagger H_2)^2+\lambda_3(H_1^\dagger H_1)(H_2^\dagger H_2)
+\lambda_4 (H_1^\dagger H_2)(H_2^\dagger H_1)\nonumber \\
&+
\frac{\lambda_5}{2}(H_1^\dagger H_2)^2+\left\{
\lambda_6 (H_1^\dagger H_1)+\lambda_7 (H_2^\dagger H_2)\right\} (H_1^\dagger H_2)+{\rm h.c.}.
\end{align}
In the basis shown in Eq. (\ref{HiggsBasis}), the Higgs boson masses can be described by the dimensionless parameters and $M_{22}$ using the stationary conditions for the Higgs doublets:
\begin{align}
m_{H^+}^2&=M_{22}^2+\frac{v^2}{2}\lambda_3, \nonumber \\
m_A^2-m_{H^+}^2&=-\frac{v^2}{2}(\lambda_5-\lambda_4),\nonumber \\
(m_H^2-m_h^2)^2&=\left\{m_A^2+(\lambda_5-\lambda_1)v^2\right\}^2+4\lambda_6^2v^4,\nonumber \\
\sin 2\theta_{\beta\alpha}&=-\frac{2\lambda_6 v^2}{m_H^2-m_h^2}.
\end{align}
Especially, when $c_{\beta \alpha}$ is close to zero (that is, $\lambda_6 \sim 0$), we obtain the following
simple expressions for the Higgs boson masses:
\begin{align}
  m_h^2& \simeq \lambda_1 v^2,\nonumber \\
  m_H^2& \simeq m_A^2+\lambda_5 v^2,\nonumber \\
  m_{H^+}^2 &= m_A^2-\frac{\lambda_4-\lambda_5}{2} v^2,\nonumber \\
  m_A^2 & = M_{22}^2+\frac{\lambda_3+\lambda_4-\lambda_5}{2} v^2.
  \label{Higgs_spectrum2}
\end{align}
Note that fixing the couplings, $\lambda_3$, $\lambda_4$ and $\lambda_5$, the heavy Higgs boson masses
are expressed by the CP-odd Higgs boson mass $m_A$, which we treat as a free
parameter of the model.
The contribution to the Peskin-Takeuchi T-parameter \cite{peskin} should be taken into account, so that
we assume that it is suppressed by the
degeneracy between $m_A$ and $m_{H^+}$ as well as the small Higgs mixing parameter $c_{\beta \alpha}$.
Therefore, we set $\lambda_4=\lambda_5$ in our analysis, which corresponds to
$m_A=m_{H^+}$.

%%%%%%%%%%%%%%%%%%%%%%%%%%%%%%%%%

\section{Solution to the CMS excess in $h\rightarrow \mu\tau$ and the muon g-2 anomaly}
The CMS collaboration has reported an excess in a Higgs boson decay mode
$h\rightarrow \mu\tau$.
Furthermore, it is known that there is a discrepancy between the measured value and the SM prediction of the muon anomalous magnetic moment (muon g-2). The both anomalies cannot be explained by the SM, and hence
they might be an indication of physics beyond the SM.
In this section, we discuss whether the general 2HDM can accommodate both anomalies simultaneously,
and investigate the parameter space where both anomalies can be explained.

%%%%%%%%%%%%%%%%%%%%%%%%%%%%%%%%%%%%%%%%%%%%%%%%%%%%%%%%%%%%%%%%%%%%%%
\subsection{$h\rightarrow \mu\tau$}
An excess in $h\rightarrow \mu\tau$ decay mode has been reported by the CMS collaboration: The best fit value of the branching ratio is
%{\color{red}
 $ {\rm BR}(h\rightarrow \mu \tau)=(0.84^{+0.39}_{-0.37}) \%$ \cite{Khachatryan:2015kon}.
%where the final state is a sum of $\mu^+\tau^-$ and $\mu^-\tau^+$.
%If this is confirmed by the ATLAS collaboration and the further data, it is an indication of
%new physics because it can not occur in the SM.
As discussed in the Introduction, the ATLAS collaboration has also shown the result,
${\rm BR}(h\rightarrow \mu \tau)=(0.77\pm{0.62}) \%$~\cite{ATLAS_hmt},
which is consistent with the CMS one. 
It would be an indication of new physics
because the SM can not accommodate the excess.
Since in the general 2HDM the SM-like
Higgs boson has flavor violating Yukawa interactions as discussed in the previous section,
the excess can be easily explained.
The expression of the branching ratio of the $h\rightarrow \mu\tau$ process is given by
\begin{align}
  {\rm BR}(h \rightarrow \mu \tau)
  &= \frac{\Gamma(h \rightarrow \mu^+\tau^-)
    +\Gamma(h \rightarrow \mu^-\tau^+)}{\Gamma_h} \nonumber \\
  &=\frac{c_{\beta\alpha}^2
    (|\rho_e^{\mu \tau}|^2+|\rho_e^{\tau\mu}|^2)m_h}{16\pi\Gamma_h},
\end{align}
where $\Gamma_h$ is a total decay width of Higgs boson $h$ and we adopt $\Gamma_h=4.1$ MeV
in this paper. In order to accommodate the CMS excess, the $\mu-\tau$ flavor violating Yukawa
couplings need to satisfy the following condition:
%{\color{red}
\begin{align}
  \bar{\rho}^{\mu\tau}&\equiv \sqrt{\frac{|\rho_e^{\mu \tau}|^2+|\rho_e^{\tau\mu}|^2}{2}}
  \nonumber \\
  &\simeq 0.26 \left(\frac{0.01}{|c_{\beta\alpha}|}  \right)
  \sqrt{\frac{{\rm BR}(h\rightarrow \mu \tau)}{0.84\times 10^{-2}}}.
\end{align}
%}
It is interesting to note that even in the small Higgs mixing $(|c_{\beta \alpha}|\simeq 0.01)$,
the $\mu-\tau$ flavor violating Yukawa couplings with the order of 0.1 can achieve the CMS excess.
%%%%%%%%%%%%%%%%%%%%%%%%%%%%%%%%%%%%%%%%%%%%%%%%%%%%%%%%%%%%%%%%%%%%%%%%%%%%%%%%%%%%%%%%%%%%%%%%%%%%
\subsection{ The muon anomalous magnetic moment (muon g-2)}
We have shown that the $\mu-\tau$ flavor violating Yukawa couplings
in the general 2HDM 
explain the CMS excess in the $h\rightarrow \mu\tau$ decay mode.
Here we consider extra contributions to the muon anomalous magnetic moment (muon g-2)
generated by the $\mu-\tau$ flavor violating Yukawa couplings.

\begin{figure}[h]
  \begin{center}
    %    {\epsfig{figure=FIGs/muonG2.pdf,width=0.6\textwidth}}
    \includegraphics[width=0.6\textwidth]{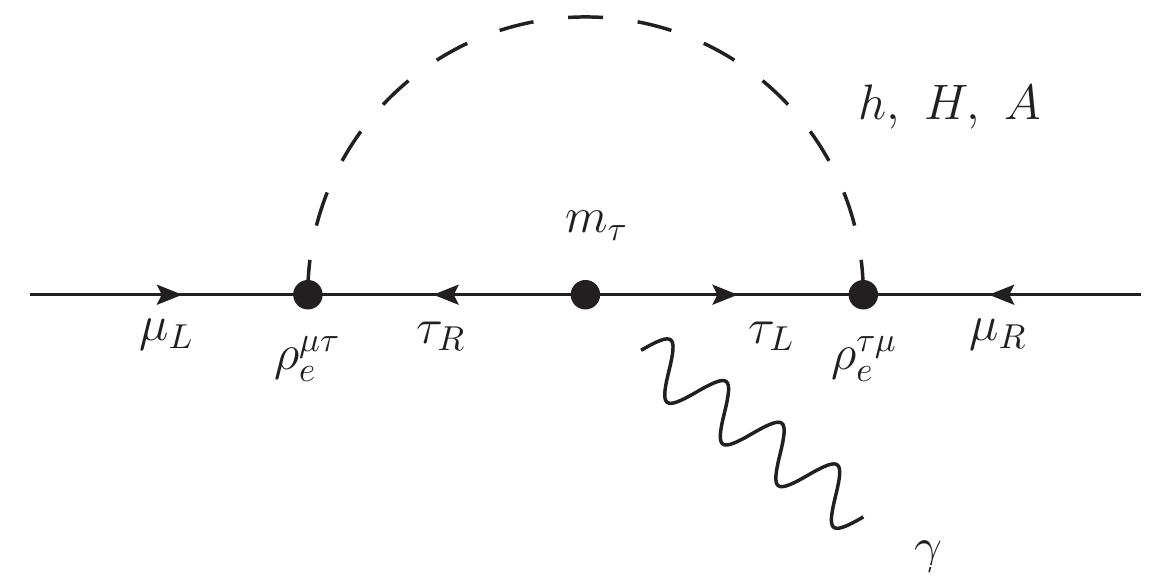}
    \caption{A Feynman diagram for neutral Higgs boson contributions to the muon g-2.
      A photon is attached somewhere in the charged lepton line.}
    \label{muonG2}
  \end{center}
\end{figure}

A discrepancy between the measured value ($a_\mu^{\rm Exp}$) and the standard model prediction
($a_\mu^{\rm SM}$) of the muon g-2 has been reported~\cite{Agashe:2014kda}. For example, Ref.~\cite{Hagiwara:2011af}
shows the following result:
\begin{eqnarray}
  a_{\mu}^{\rm Exp}-a_\mu^{\rm SM}=(26.1\pm 8.0)\times 10^{-10}.
  \label{muonG2_anomaly}
\end{eqnarray}
Here we consider whether the extra contributions induced by the $\mu-\tau$
flavor violating interactions can accommodate the muon g-2 anomaly.
The effective operator for the muon g-2 is expressed by
\begin{eqnarray}
{\cal L}=\frac{v}{\Lambda^2}\bar{\mu}_L\sigma^{\mu\nu} \mu_R F_{\mu\nu}+{\rm h.c.}.
\end{eqnarray}
We note that the chirality of muon is flipped in the operator. Therefore,
if there is a large chirality flip induced by the new physics,
it can enhance the extra contributions to the muon g-2~\cite{Kanemitsu:2012dc}.
A Feynman diagram for the extra contributions of the neutral Higgs bosons to the muon g-2,
induced by the $\mu-\tau$ flavor violating Yukawa couplings, is shown in Fig.~\ref{muonG2}.
As shown in Fig.~\ref{muonG2}, the chirality flip
occurs in the internal line of $\tau$ lepton in the diagram. Therefore, it induces
the $O(m_\tau/m_\mu)$ enhancement in the extra contributions to the muon g-2, compared with
the one generated by the flavor-diagonal Yukawa coupling. We stress that the $\mu-\tau$
flavor violating interaction is essential to obtain such an enhancement.
Note that both
couplings $\rho_e^{\mu\tau}$ and $\rho_e^{\tau\mu}$ should be nonzero to get the chirality flip in the internal $\tau$ lepton line.
The expression of the enhanced extra contribution $\delta a_\mu$ is given by
\begin{align}
  \delta a_\mu&=\frac{m_\mu m_\tau \rho_e^{\mu\tau}\rho_e^{\tau\mu}}{16\pi^2}
  \left[\frac{c^2_{\beta\alpha}(\log\frac{m_h^2}{m_\tau^2}-\frac{3}{2})}{m_h^2}
%    \right.
%    \nonumber \\
%    &\quad\left.
    +\frac{s^2_{\beta\alpha}(\log\frac{m_H^2}{m_\tau^2}-\frac{3}{2})}{m_H^2} 
-\frac{\log\frac{m_A^2}{m_\tau^2}-\frac{3}{2}}{m_A^2}
\right],
  \label{a_mu}
\end{align}
where we have assumed that $\rho_e^{\mu \tau} \rho_e^{\tau\mu}$ is real, for simplicity.
We will discuss the effect of an imaginary part of these Yukawa couplings later.
We note that a degeneracy of all neutral Higgs bosons suppresses the extra contribution
to the muon g-2, as seen in Eq.~(\ref{a_mu}).

\begin{figure}[h]
  \begin{center}
    \includegraphics[clip,width=0.7\textwidth]{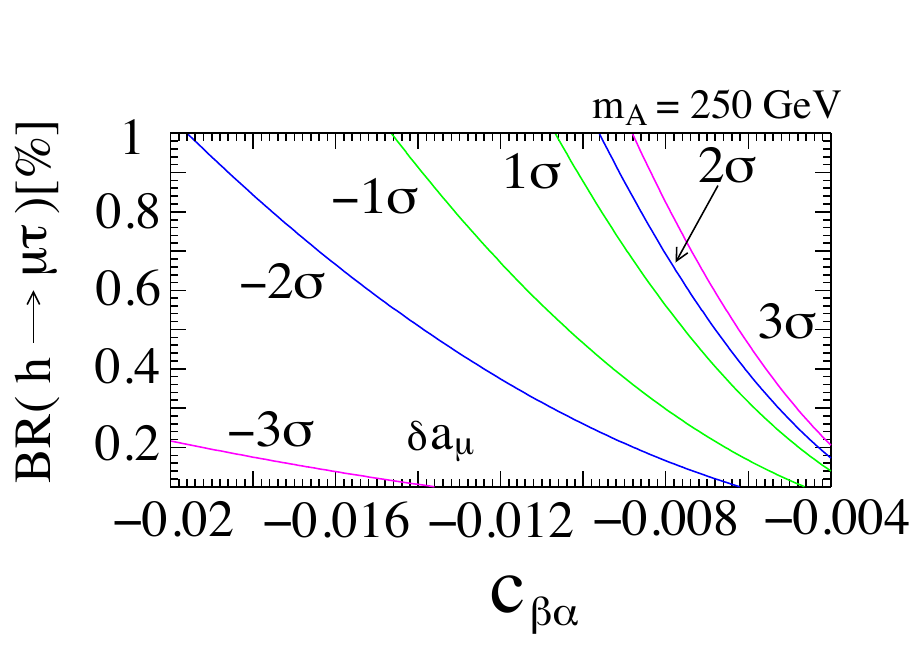}
    \includegraphics[clip,width=0.7\textwidth]{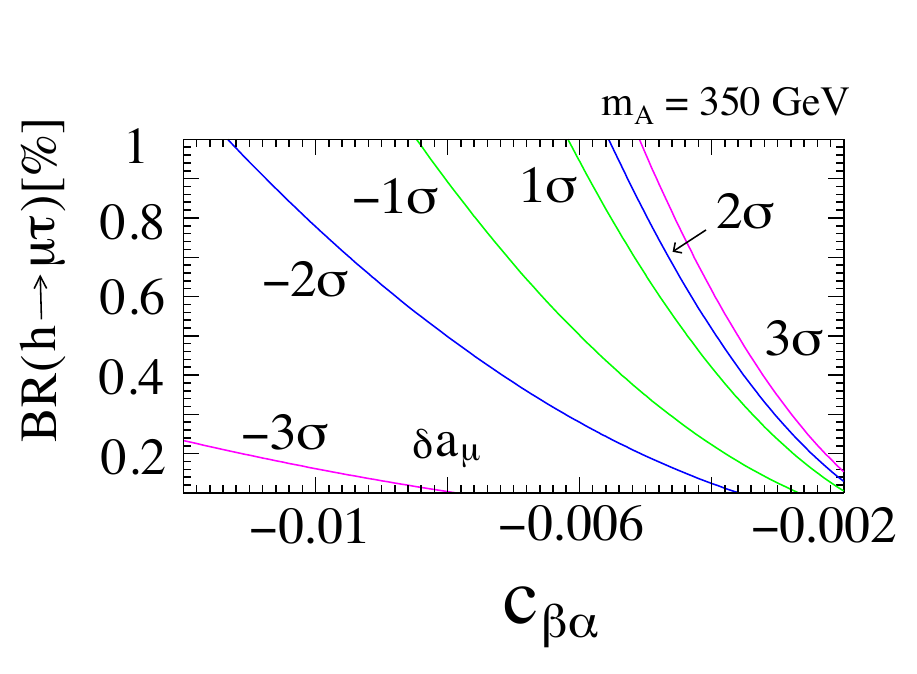}
    \caption{Numerical result for $\delta a_\mu$ as a function of $c_{\beta\alpha}$
      and ${\rm BR}(h\rightarrow \mu\tau)$ for $m_A=250$ GeV (upper figure) and 350 GeV (lower figure).
      Regions where the muon g-2 anomaly in Eq.~(\ref{muonG2_anomaly}) is explained within $\pm 1\sigma$,
      $\pm 2\sigma$ and $\pm 3\sigma$ are shown.
      Here we determine the mass spectrum of heavy Higgs bosons assuming $\lambda_4=\lambda_5=0.5$
      in Eq.~(\ref{Higgs_spectrum2}).
      We have assumed $\rho_e^{\mu\tau}\rho_e^{\tau\mu}<0$ with $\rho_e^{\mu\tau}=-\rho_e^{\tau\mu}$
      to obtain the positive contribution to $\delta a_\mu$.    
    }
    \label{muonG2_vs_htm}
  \end{center}
\end{figure}
In Fig.~\ref{muonG2_vs_htm}, we show numerical results for the extra contribution
to muon g-2 ($\delta a_\mu$) as a function
of $c_{\beta\alpha}$ and ${\rm BR}(h\rightarrow \mu\tau)$ for $m_A=250$ GeV (upper figure) and
350 GeV (lower figure).
Regions where the muon g-2 anomaly in
Eq.~(\ref{muonG2_anomaly}) is explained within $\pm 1\sigma$, $\pm 2\sigma$ and $\pm 3\sigma$ are shown.
Here we have fixed the mass spectrum of heavy Higgs bosons assuming $\lambda_4=\lambda_5=0.5$
in Eq.~(\ref{Higgs_spectrum2}). We have assumed that $\rho_e^{\mu\tau}\rho_e^{\tau\mu}<0$ with
$\rho_e^{\mu\tau}=-\rho_e^{\tau\mu}$ to obtain the positive contribution to $\delta a_\mu$.
We only discuss the case with $c_{\beta\alpha}<0$, however, the predictions of $\delta a_\mu$ and
${\rm BR}(h \rightarrow \mu\tau)$ do not change even if the sign of $c_{\beta\alpha}$
is flipped ($c_{\beta\alpha}\rightarrow -c_{\beta\alpha}$).
One can see that there are regions where both anomalies of the muon g-2 and $h \rightarrow \mu\tau$
can be explained in the 2HDM. We note that although larger ${\rm BR}(h \rightarrow \mu \tau)$ is
preferred by the muon g-2
anomaly, the regions where ${\rm BR}(h \rightarrow \mu \tau)$ is smaller than the one suggested by
the CMS result are also allowed by the muon g-2 anomaly if $|c_{\beta\alpha}|$ is small.

\begin{figure}[h]
  \begin{center}
%    {\epsfig{figure=note_FIG/amu_case1_no2.eps,width=0.8\textwidth}}
    \includegraphics[width=0.7\textwidth]{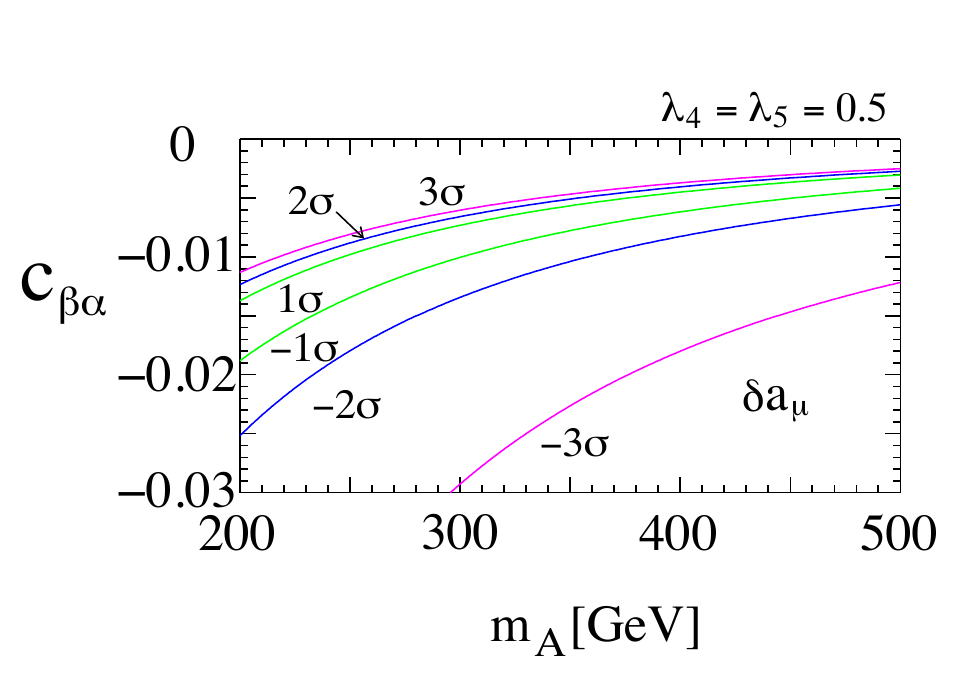}
%  \begin{center}
%    {\epsfig{figure=note_FIG/amu_case1_nega_no2.eps,width=0.6\textwidth}}
%  \end{center}
  \caption{Numerical result for $\delta a_\mu$ as a function of $m_A$ and
    $c_{\beta\alpha}$ assuming ${\rm BR}(h\rightarrow \mu\tau)=0.84\%$.
    Regions which explain the muon g-2 anomaly in Eq.~(\ref{muonG2_anomaly}) within $\pm 1\sigma$,
    $\pm 2\sigma$ and $\pm 3\sigma$ are shown.
    Here we determine the mass spectrum of heavy Higgs bosons as a function of $m_A$
    assuming $\lambda_4=\lambda_5=0.5$ in Eq.~(\ref{Higgs_spectrum2}).
    We have assumed $\rho_e^{\mu\tau}=-\rho_e^{\tau\mu}$ to obtain the positive contribution
    to $\delta a_\mu$ and the Yukawa couplings
    $\rho_e^{\mu\tau(\tau\mu)}$ are fixed to realize ${\rm BR}(h\rightarrow \mu\tau)=0.84\%$.
  }
  \label{para_muonG2}
  \end{center}
\end{figure}

In Fig.~\ref{para_muonG2}, the numerical result for the $\delta a_\mu$
is shown as a function of $m_A$ and $c_{\beta\alpha}$ fixing that
${\rm BR}(h\rightarrow \mu\tau)=0.84\%$.
Regions that explain the muon g-2 anomaly in Eq.~(\ref{muonG2_anomaly}) within
$\pm 1 \sigma$, $\pm 2 \sigma$ and $\pm 3 \sigma$ are shown.
Here we take $\lambda_4=\lambda_5=0.5$ to determine the mass spectrum of heavy Higgs bosons
(shown in Eq.~(\ref{Higgs_spectrum2})) as a function of $m_A$. We assume that the
Yukawa couplings $\rho_e^{\tau\mu}$ and $\rho_e^{\mu\tau}$ are determined to realize 
${\rm BR}(h\rightarrow \mu\tau)=0.84\%$ with $\rho_e^{\mu\tau}=-\rho_e^{\tau\mu}$.
When $|c_{\beta\alpha}|$ gets smaller, $\delta a_\mu$ increases because the Yukawa
couplings $\rho_e^{\mu\tau(\tau\mu)}$ become larger with the fixed ${\rm BR}(h\rightarrow \mu\tau)$.
It is interesting to see that the 2HDM can explain both anomalies of the muon g-2 and
$h\rightarrow \mu\tau$ when $|c_{\beta\alpha}|$ is small ($|c_{\beta\alpha}|\sim 0.01$)
for $m_A=200-500$ GeV. We note that the small mixing $|c_{\beta\alpha}|$ is consistent with the current
results of the Higgs coupling measurements as well as the constraints from the electroweak
observables.
\begin{figure}[h]
  \begin{center}
    %    {\epsfig{figure=note_FIG/amu_negaC0008.eps,width=0.8\textwidth}}
    \includegraphics[width=0.7\textwidth]{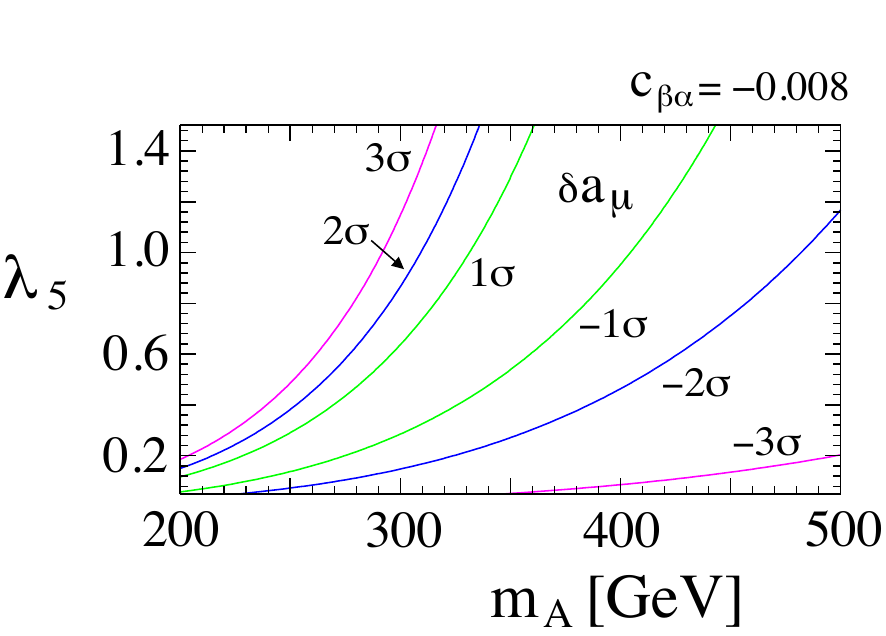}
  \end{center}
  \caption{Numerical result for $\delta a_\mu$ as a function of $m_A$ and $\lambda_5$
    assuming ${\rm BR}(h\rightarrow \mu\tau)=0.84\%$ with the fixed $c_{\beta\alpha}~(=-0.008)$.
    Here we have assumed $\lambda_4=\lambda_5$ and $\rho_e^{\mu\tau}=-\rho_e^{\tau\mu}$.}
  \label{para2_muonG2}
\end{figure}
In Fig.~\ref{para2_muonG2}, similar to Fig.~\ref{para_muonG2},
the numerical result for the $\delta a_\mu$ is shown as a function of $m_A$ and $\lambda_5$
fixing that ${\rm BR}(h\rightarrow \mu\tau)=0.84~\%$ and $c_{\beta\alpha}=-0.008$.
We have assumed that $\rho_e^{\tau\mu}=-\rho_e^{\mu\tau}$ and $\lambda_4=\lambda_5$.
As $\lambda_5$ gets larger, the $\delta a_\mu$ becomes larger because
the non-degeneracy between $H$ and $A$ increases and it enhances the $\delta a_\mu$.
Figs.~\ref{muonG2_vs_htm}, \ref{para_muonG2} and \ref{para2_muonG2} show the
typical interesting regions
which explain both anomalies of the muon g-2 and $h\rightarrow \mu\tau$.

%%%%%%%%%%%%%%%%%%%%%%%%%%%%%%%%%%%%%
%%%%%%%%%%%%%%%%%%%%%%%%%%%%%%%%%%%%%

\section{$\tau$- and $\mu$-physics in this scenario}
So far, we have seen that the general 2HDM with the $\mu-\tau$ flavor violation accommodates
the muon g-2 anomaly and the CMS excess in $h\rightarrow \mu\tau$ decay, simultaneously.
The parameter regions
with $|c_{\beta\alpha}|\sim 0.01$ and $m_A\sim O(100)$ GeV are typically interesting. In this section,
we investigate what kinds of predictions and/or constraints in $\tau$- and $\mu$-physics
we have in this scenario.
\subsection{$\tau \rightarrow \mu \gamma$}
\begin{figure}[h]
  \includegraphics[width=0.35\textwidth,angle=90]{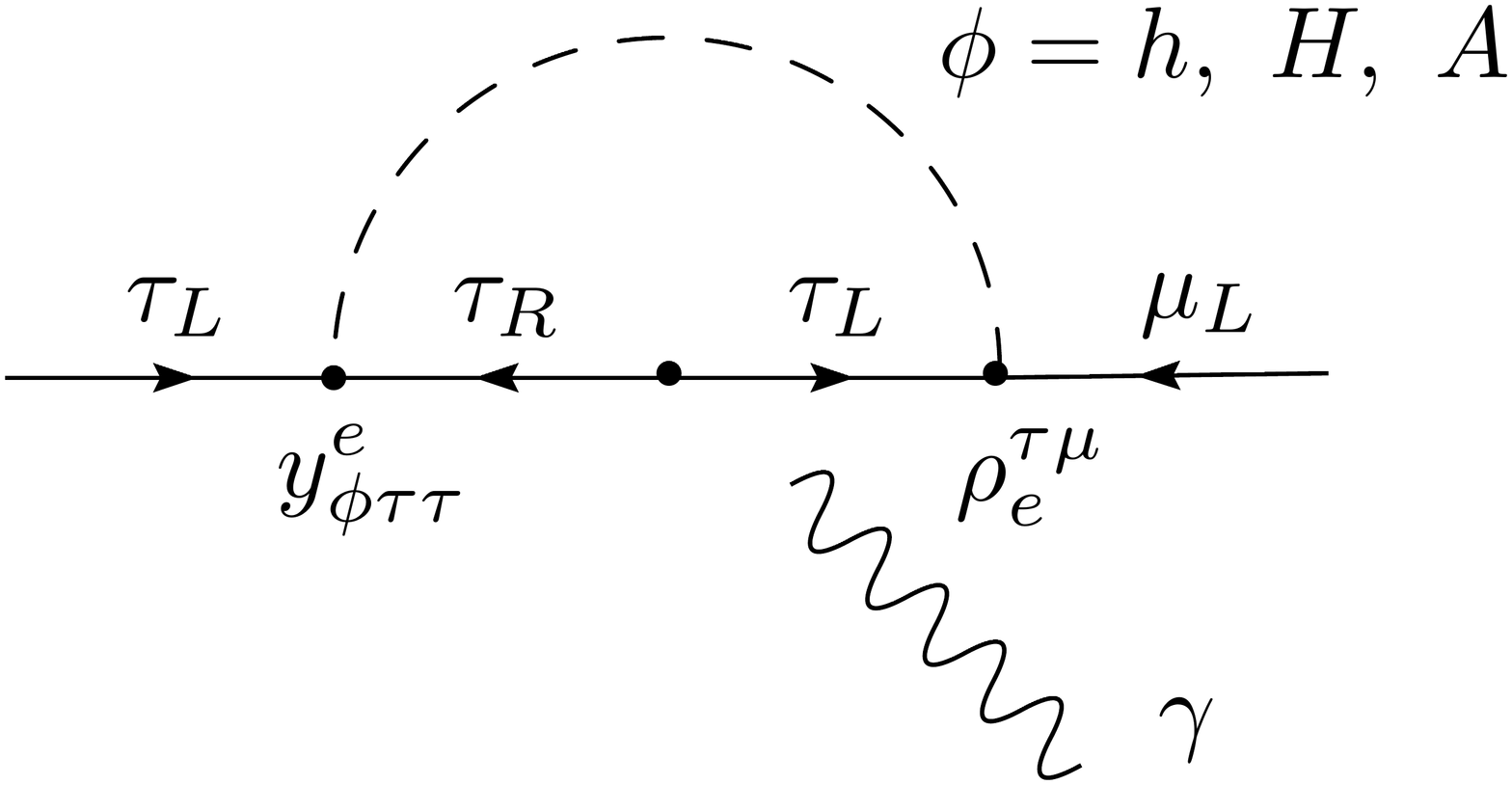}
  \includegraphics[width=0.49\textwidth,angle=0]{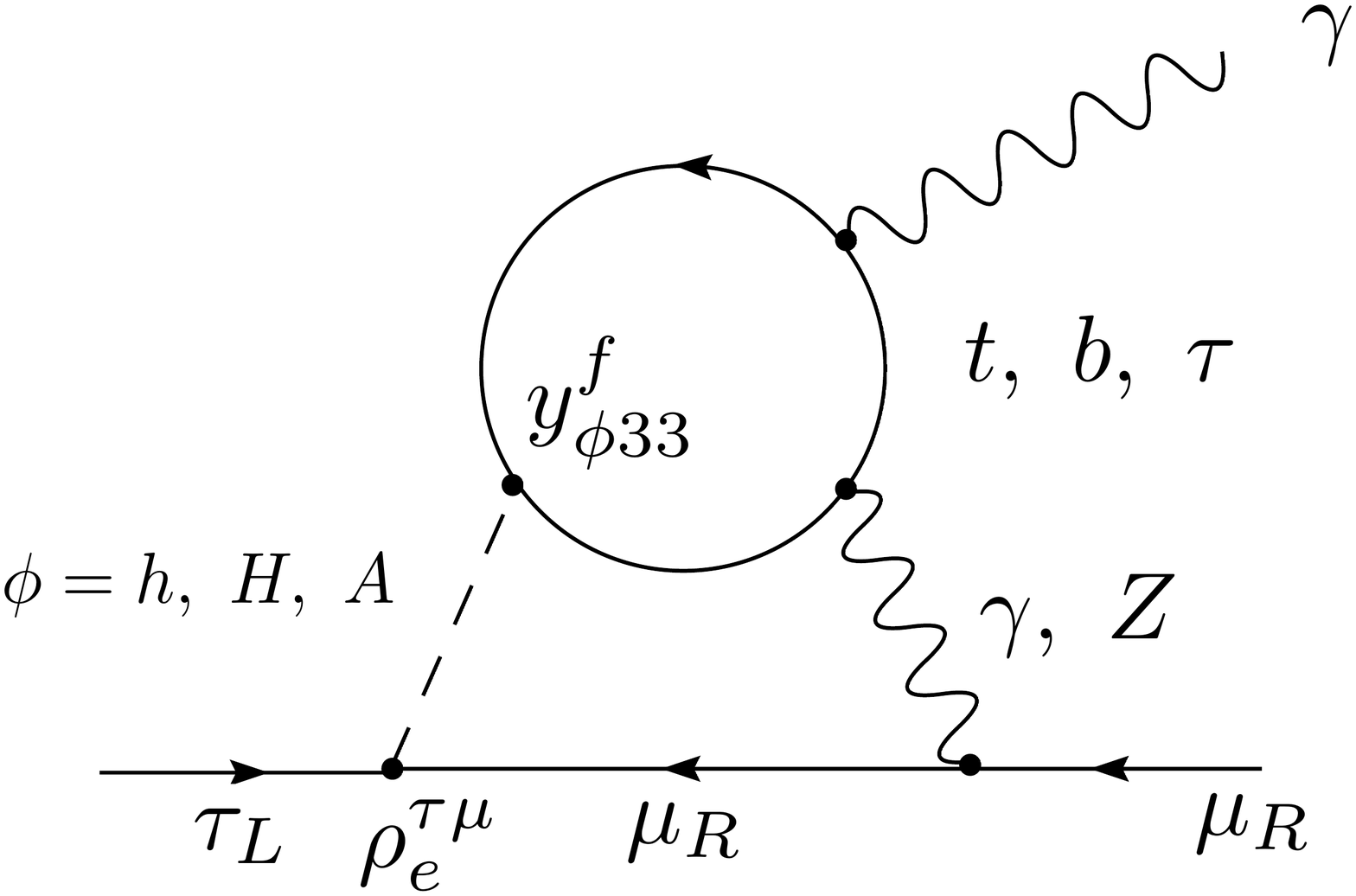}
  \caption{Some of Feynman diagrams which contribute to $\tau \rightarrow \mu\gamma$ processes
    at one-loop level (left figure) which are induced by $y^e_{\phi\; \tau\tau}~(\phi=h,~H,~A)$
    Yukawa couplings and at two-loop level (right figure) which are Barr-Zee type
    contributions and induced by
    the third generation fermions via $y_{\phi\; 33}^f~(f=u,~d,~e)$ Yukawa couplings.
    Diagrams where the fermion chiralities are flipped also contribute.
    We also have a Barr-Zee type two-loop contribution induced by W-loop, which
    is not shown here.
  }
  \label{tmg_diagrams}
\end{figure}

The first process we would like to discuss is $\tau \rightarrow \mu\gamma$.
The $\mu-\tau$ flavor violating Yukawa couplings induce the flavor violating phenomena $\tau \rightarrow \mu\gamma$, as shown, for example, in Fig.~\ref{tmg_diagrams}.
We parametrize the decay amplitude ($T_{\tau \rightarrow \mu \gamma}$) as follows:
\begin{eqnarray}
  T_{\tau \rightarrow \mu \gamma}
  =e \epsilon^{\alpha *}(q)\bar{u}_\mu(p-q) m_\tau i\sigma_{\alpha \beta} q^\beta (A_L P_L+A_R P_R)u_\tau(p),
\end{eqnarray}
where $P_{R,~L}(=(1\pm \gamma_5)/2)$ are chirality projection operators, and $e,~\epsilon^\alpha,~q,~p$
and $u_f$ are the electric charge, a photon polarization vector, a photon momentum, a $\tau$ momentum,
and a spinor of the fermion $f$, respectively. The branching ratio is given by
\begin{eqnarray}
  \frac{{\rm BR}(\tau \rightarrow \mu \gamma)}{{\rm BR}(\tau \rightarrow \mu \bar{\nu}_\mu \nu_\tau)}
  =\frac{48\pi^3\alpha\left(|A_L|^2+|A_R|^2\right)}
  {G_F^2},
\end{eqnarray}
where $\alpha$ and $G_F$ are the fine structure constant and the Fermi constant, respectively.
The lepton flavor violating Higgs contributions to $A_L$ and $A_R$ via $y^e_{\phi\; \tau\tau}~(\phi=h,~H,~A)$ Yukawa interactions
at one-loop level (left figure in Fig.~\ref{tmg_diagrams}) are given by~\footnote{
  Yukawa couplings $y^e_{\phi\; \mu \mu}$ also contribute to $\tau\rightarrow \mu\gamma$. However, the SM
  part of $y^e_{\phi\; \mu\mu}$ is smaller than the one of $y^e_{\phi\; \tau\tau}$, and $\rho_e^{\mu\mu}$ is strongly constrained by
  $\tau\rightarrow 3\mu$ process as discussed later. Therefore we have neglected the contributions
  from $y^e_{\phi\;\mu\mu}$.}
\begin{align}
  A_{L,~R}&=\sum_{\phi=h,~H,~A,~H^-} A_{L,~R}^{\phi},\\
  A_L^{\phi}&= \frac{y^{e*}_{\phi\;\tau \mu}}{16\pi^2 m_\phi^2}
  \left[y^{e*}_{\phi\;\tau\tau}\left(\log\frac{m_\phi^2}{m_\tau^2}-\frac{3}{2}\right)
    +\frac{y^e_{\phi\;\tau\tau}}{6}\right],~~(\phi=h,~H,~A)\nonumber \\
  A_R^{\phi}&=A_L^{\phi}|_{y^{e*}_{\phi\; \tau\mu}\rightarrow y^e_{\phi\; \mu\tau},
    ~~y^{e}_{\phi \;\tau \tau} \leftrightarrow y^{e*}_{\phi\; \tau\tau}},~~
  (\phi=h,~H,~A),\nonumber \\
  A_L^{H^-}&= -\frac{(\rho_e^\dagger \rho_e)^{\mu \tau}}{192\pi^2 m_{H^-}^2},~~
  A_R^{H^-}=0,
\end{align}
where $A_{L,~R}^{\phi}~(\phi=h,~H,~A,~H^-)$ are the $\phi$ contributions at the one loop level.
The Yukawa couplings $y^e_{\phi\; \tau\tau}~(\phi=h,~H,~A)$ are given in Eq.~(\ref{yukawa}).
Here we have neglected the $O(m_\mu/m_\tau)$ contributions.
%%%%%%%%%%%%%%%%###################

We also find that the Barr-Zee type contributions ($A_{L,R}^{\rm BZ}$) at the two loop level are important
and dominant in the most of cases. The third generation fermion contributions via $y^f_{\phi\; 33}~(f=u,~d,~e)$
Yukawa couplings\footnote{In our notation, $y_{\phi\; 33}^u=y_{\phi\; tt}^u,~y_{\phi\; 33}^d=y_{\phi\; bb}^d,
~y_{\phi\; 33}^e=y_{\phi\; \tau\tau}^e$.} (shown in the right figure in Fig.~\ref{tmg_diagrams}) and the W-boson
contribution (not shown in Fig.~\ref{tmg_diagrams}) are given by
\footnote{The Barr-Zee contributions to $\mu \rightarrow e\gamma$ have been studied in
  Ref.~\cite{Chang:1993kw}. The application to $\tau \rightarrow \mu\gamma$ is apparent
  and we adopt their results for $\tau \rightarrow \mu\gamma$.}
\begin{align}
  A^{\rm BZ}_{L}&=-\sum_{\phi=h,H,A;f=u,d,e}
  \frac{N_C Q_f \alpha}{8\pi^3}
  \frac{y^{e*}_{\phi\;\tau\mu}}{m_\tau m_{f_3}}
%  \frac{y^f_{\phi 33}y^e_{\phi\tau\mu}
%  \nonumber \\
%  &\times
  \left[
  Q_f\left\{
    {\rm Re} (y^f_{\phi\; 33}) 
    F_H\left(x_{f\phi}\right) 
  - i {\rm Im} (y^f_{\phi\; 33})
  F_A\left(x_{f\phi}\right)\right\}\right.
  \nonumber \\
  &\left. +\frac{(1-4 s_W^2)(2T_{3f}-4Q_f s_W^2)}{16s_W^2 c_W^2}
  \left\{
    {\rm Re} (y^f_{\phi\; 33}) 
    \tilde{F}_H\left(x_{f\phi},x_{fZ}\right) 
  - i {\rm Im} (y^f_{\phi\; 33})
  \tilde{F}_A\left(x_{f\phi},x_{f Z}\right)\right\}\right]
  \nonumber \\  
  &+\sum_{\phi=h,H}\frac{\alpha}{16\pi^3}\frac{g_{\phi WW} y^{e*}_{\phi\;\tau\mu}}{m_\tau v}
  \left[
    3F_H\left(x_{W\phi}\right)
    +\frac{23}{4} F_A\left(x_{W\phi}\right)
    \nonumber \right.\\
    &\hspace{4cm}    +\frac{3}{4} G\left(x_{W\phi}\right) 
    +\frac{m_\phi^2}{2 m_W^2}\left\{
    F_H\left(x_{W\phi}\right)-F_A\left(x_{W\phi}\right)
    \right\}\nonumber \\
    &+\frac{1-4s_W^2}{8s_W^2}\left\{\left(
    5-t_W^2+\frac{1-t_W^2}{2 x_{W\phi}}\right) \tilde{F}_H (x_{W\phi}, x_{WZ})
    \right.\nonumber \\
    &\left.\left.+\left(7-3t_W^2 -\frac{1-t_W^2}{2x_{W\phi}}\right) \tilde{F}_A(x_{W\phi}, x_{WZ})
    +\frac{3}{2}\left\{F_A(x_{W\phi})+G(x_{W\phi})\right\} \right\}   \right],  \label{Barr-Zee}
\\
  A_R^{\rm BZ} &=A_L^{\rm BZ}(y^{e*}_{\phi\;\tau\mu}\rightarrow
  y^{e}_{\phi\;\mu\tau},~i\rightarrow -i),
\end{align}
where $x_{f\phi}=m_{f_3}^2/m_\phi^2$, $x_{fZ}=m_{f_3}^2/m_Z^2~(f_3=t,b,\tau~{\rm for}~f=u,d,e)$,
$x_{W\phi}=m_W^2/m_\phi^2$ and $x_{WZ}=m_W^2/m_Z^2$,
and $s_W^2=\sin\theta_W^2,~c_W^2=\cos\theta_W^2$ and $t_W^2=\tan\theta_W^2$.
$T_{3f}$ denotes the isospin of the fermion.
Here the couplings $g_{\phi WW}=s_{\beta\alpha}~(c_{\beta\alpha})$ for $\phi=h$ ($\phi=H$).
Functions $F_{H,~A}$, $G$ and $\tilde{F}_{H,~A}$ are defined by
\begin{align}
  F_{H}(z)&=\frac{z}{2}\int_0^1 dx \frac{1-2x(1-x)}{x(1-x)-z}\log \frac{x(1-x)}{z},\\
  F_A(z) &=\frac{z}{2}\int_0^1 dx \frac{1}{x(1-x)-z}\log \frac{x(1-x)}{z},\\
  G(z)&=-\frac{z}{2}\int_0^1 dx \frac{1}{x(1-x)-z}\left[
    1-\frac{z}{x(1-x)-z}\log \frac{x(1-x)}{z}
    \right],\\
  \tilde{F}_H(x,y)&=\frac{xF_H(y)-yF_H(x)}{x-y},\\
  \tilde{F}_A(x,y)&=\frac{xF_A(y)-yF_A(x)}{x-y}.
\end{align}
Note if the Yukawa couplings $\rho_f^{ij}$ are real, ${\rm Im}(y_{\phi\; 33}^f)=0$ for $f=h~{\rm and}~H$, and
${\rm Re} (y^{f}_{\phi\; 33})=0$ for $f=A$ are satisfied, as shown in Eq.~(\ref{yukawa}).
For simplicity, we assume that all $\rho_f^{ij}$ are real
in the calculation of $\tau \rightarrow \mu \gamma$.
The contribution in the first line (the second line) of Eq.~(\ref{Barr-Zee})
comes from the effective $\phi \gamma\gamma$ vertex ($\phi Z\gamma$ vertex)
induced by the third generation fermion loop, and the one in the third and the forth lines
(the fifth and sixth lines) originates from the effective $\phi \gamma\gamma$ vertex
($\phi Z\gamma$ vertex) generated by the W-boson loop.
In the analysis of $\tau \rightarrow \mu \gamma$ in
Ref.~\cite{Omura:2015nja}, we have not included the Barr-Zee type contributions induced by the effective
$\phi \gamma Z$ vertex since they are sub-dominant contributions. Here we include them and find they
change the results by about $10\%$.

\begin{figure}[h]
  \begin{center}
    \includegraphics[clip,width=0.65\textwidth]{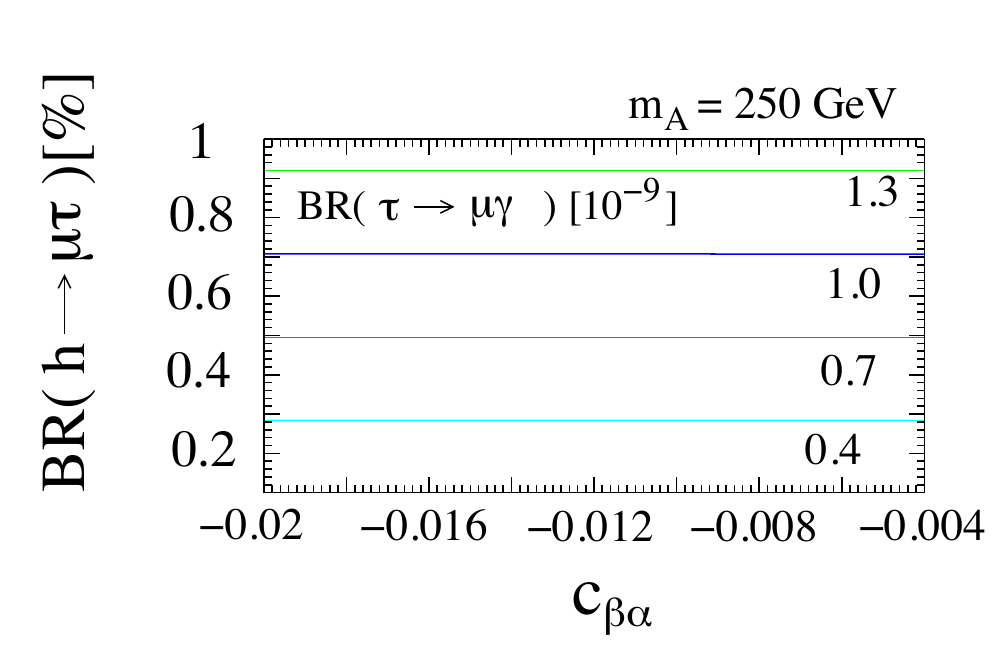}
    \includegraphics[clip,width=0.65\textwidth]{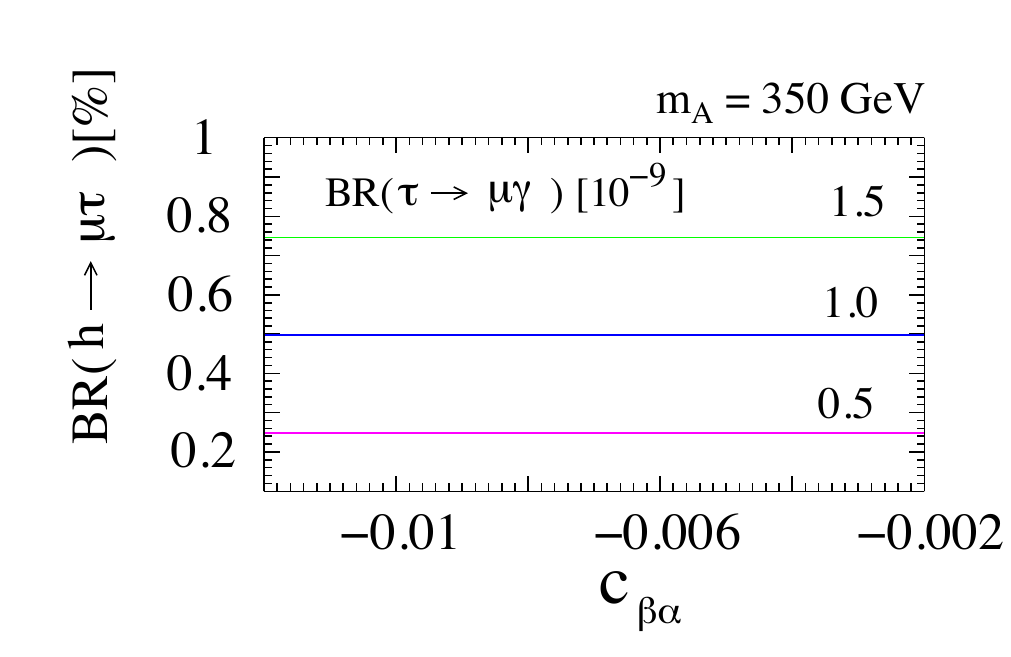}
    \caption{Numerical result for ${\rm BR}(\tau \rightarrow \mu\gamma)$ as a function of $c_{\beta\alpha}$
      and ${\rm BR}(h\rightarrow \mu\tau)$ in the same parameter set of Fig.~\ref{muonG2_vs_htm}.
      Lines for ${\rm BR}(\tau\rightarrow \mu\gamma)/10^{-9}=0.4,~0.7,~1.0$ and $1.3$
      ($0.5,~1.0$ and $1.5$) are shown for $m_A=250$ GeV ($m_A=350$ GeV).
      Here we have assumed that the extra Yukawa couplings $\rho_f$ other than
      $\rho_e^{\mu\tau~(\tau\mu)}$ are negligible.
    }
    \label{tmg_vs_htm}
  \end{center}
\end{figure}

The total amplitude $A_{L,~R}$ is a sum of all contributions,
\begin{align}
A_{L,~R}=\sum_{\phi=h,H,A,H^-}A_{L,~R}^\phi+A_{L,~R}^{\rm BZ}.
\end{align}
In Fig.~\ref{tmg_vs_htm}, numerical results for ${\rm BR}(\tau\rightarrow \mu\gamma)$ as a function of
$c_{\beta\alpha}$ and ${\rm BR}(h \rightarrow \mu\tau)$ are shown. Here the same parameter set
as the one in Fig.~\ref{muonG2_vs_htm} is taken. We have assumed that the extra Yukawa couplings $\rho_f^{ij}$
other than $\rho_e^{\mu\tau(\tau\mu)}$ are negligible.
Lines for ${\rm BR}(\tau\rightarrow \mu\gamma)/10^{-9}=0.4,~0.7,~1.0$ and
$1.3$ (upper figure)  and for ${\rm BR}(\tau\rightarrow \mu\gamma)/10^{-9}=0.5,~1.0$ 
and $1.5$ (lower figure) are shown for $m_A=250$ GeV and $350$ GeV, respectively.
As one can see, if ${\rm BR}(h\rightarrow \mu\tau)=0.84\%$ as suggested by the CMS experiment,
the branching ratio for $\tau\rightarrow \mu\gamma$ can be larger than $10^{-9}$, which might
be within the reach of the future B-factory experiment, the Belle II.
Note that the branching ratio ${\rm BR}(\tau \rightarrow \mu\gamma)$ almost does not
depend on the Higgs mixing parameter $c_{\beta \alpha}$ when ${\rm BR}(h\rightarrow \mu\tau)$
is fixed and $c_{\beta \alpha}$ is small. We also note that the cancellation between the one-loop and two-loop Barr-Zee type
contributions happens, and hence the branching ratio ${\rm BR}(\tau\rightarrow \mu \gamma)$
is not simply suppressed by the heavy Higgs boson masses.

\begin{figure}[h]
  \begin{center}
    %    {\epsfig{figure=note_FIG/BR_tmg_ext.eps,width=0.8\textwidth}}
    \includegraphics[width=0.7\textwidth]{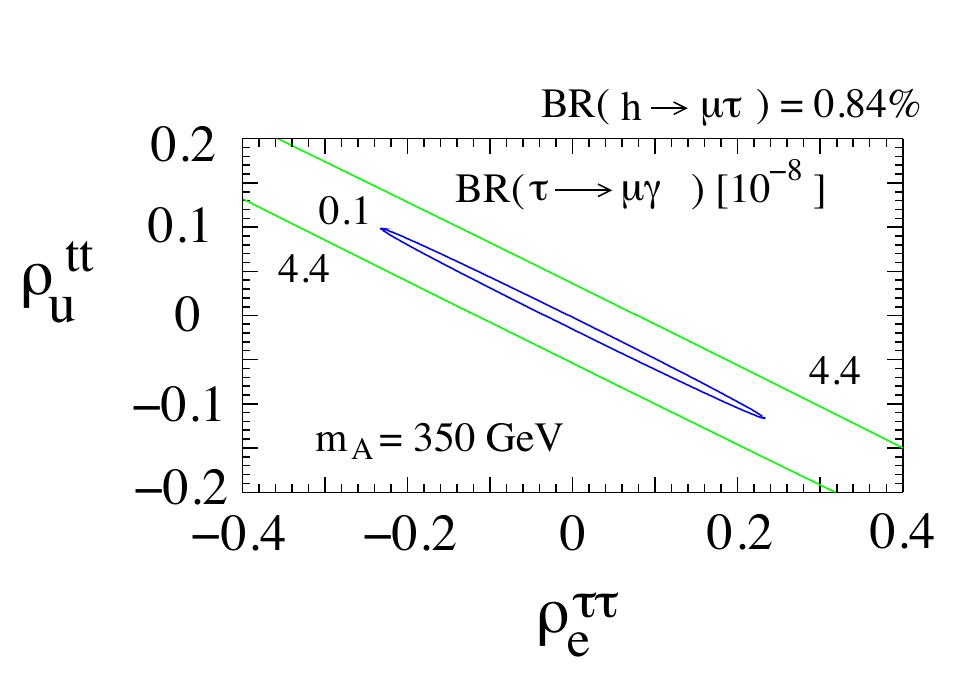}
    \caption{Numerical result for ${\rm BR}(\tau \rightarrow \mu\gamma)$ as a function of
      $\rho_e^{\tau \tau}$ and $\rho_u^{tt}$. Lines for ${\rm BR}(\tau\rightarrow \mu\gamma)/10^{-8}=0.1$
      and $4.4$ (current experimental limit) are shown. Here we have assumed $m_A=350$ GeV with $\lambda_4=\lambda_5=0.5$,
      $c_{\beta\alpha}=-0.007$ and ${\rm BR}(h\rightarrow \mu\tau)=0.84~\%$ with $\rho_e^{\mu\tau}=-\rho_e^{\tau\mu}$.
      We note that for this parameter set,
      $\delta a_\mu=2.2\times 10^{-9}$ which explains the muon g-2 anomaly within the $1\sigma$.
    }
    \label{tmg_extraYukawa}
  \end{center}
\end{figure}

If the extra Yukawa couplings other than $\rho_e^{\mu\tau(\tau\mu)}$ are not negligible, the branching
ratio ${\rm BR}(\tau\rightarrow \mu\gamma)$ could be further enhanced. For example, the extra Yukawa
coupling $\rho_e^{\tau\tau}$ can contribute at the one-loop level, and on the other hand, $\rho_u^{tt}$
can affect the branching ratio via the Barr-Zee type two-loop contribution.
In Fig.~\ref{tmg_extraYukawa}, numerical results for ${\rm BR}(\tau \rightarrow \mu\gamma)$ are shown
as a function of $\rho_e^{\tau \tau}$ and $\rho_u^{tt}$. Lines for ${\rm BR}(\tau\rightarrow \mu\gamma)/10^{-8}=0.1$
and $4.4$ (current experimental limit) are shown. Here we have assumed $m_A=350$ GeV with $\lambda_4=\lambda_5=0.5$,
$c_{\beta\alpha}=-0.007$ and ${\rm BR}(h\rightarrow \mu\tau)=0.84~\%$ with $\rho_e^{\mu\tau}=-\rho_e^{\tau\mu}$.
This parameter set can enhance the muon g-2 as $\delta a_\mu=2.2\times 10^{-9}$ which is within the $1\sigma$.
%\begin{figure}
%  \begin{center}
%    {\epsfig{figure=note_FIG/BR_tmg_minim.eps,width=0.8\textwidth}}
%  \end{center}
%\end{figure}
%
At present, the extra Yukawa couplings $\rho_e^{\tau\tau}$ and $\rho_u^{tt}$ can be still larger than, for example,
$O(0.1)$ with some correlation, however, the future experimental constraint would be significant for this scenario.
Therefore,
the $\tau\rightarrow \mu\gamma$ process would be important to probe the scenario.

\subsection{$\mu\rightarrow e\gamma$, $\tau \rightarrow e\gamma$, and electron g-2}
The $\mu-\tau$ flavor violating Yukawa couplings themselves do not generate $\mu\rightarrow e\gamma$. However,
together with $e-\mu$ or $e-\tau$ flavor-violation, $\mu \rightarrow e \gamma$ is induced. Since there is
a strong constraint from this process, $e-\mu$ and $e-\tau$ flavor violating couplings are strongly
constrained.

Similar to $\tau \rightarrow \mu\gamma$, we parametrize the decay amplitude
$(T_{\mu \rightarrow e\gamma})$ as
\begin{align}
  T_{\mu \rightarrow e\gamma}
  =e \epsilon^{\alpha *} \bar{u}_e m_\mu i \sigma_{\alpha \beta} q^\beta (A_L P_L+A_R P_R) u_\mu,
\end{align}
and the branching ratio is given by
\begin{align}
{\rm BR}(\mu \rightarrow e \gamma)=\frac{48\pi^3\alpha (|A_L|^2+|A_R|^2)}{G_F^2}.
\end{align}
The neutral Higgs contributions $A_{L,~R}^{\phi}$ ($\phi=h,~H,~A$) to $A_{L,~R}$ at the one-loop
are given by
\begin{align}
  A_L^{\phi}&=\frac{1}{16\pi^2}\sum_{i=\mu,\tau}\frac{y^{e*}_{\phi\; ie}}{m_\phi^2}
  \left[\frac{m_i}{m_\mu}y^{e *}_{\phi\; \mu i}\left(
    \log\frac{m_\phi^2}{m_i^2}-\frac{3}{2}\right)+\frac{y^{e}_{\phi\; i\mu}}{6}\right],\\
  A_R^{\phi}&=\frac{1}{16\pi^2}\sum_{i=\mu,\tau}\frac{y^{e}_{\phi\; e i }}{m_\phi^2}
  \left[\frac{m_i}{m_\mu}y^{e }_{\phi\; i \mu}\left(
    \log\frac{m_\phi^2}{m_i^2}-\frac{3}{2}\right)+\frac{y^{e *}_{\phi\; \mu i}}{6}\right],
  \label{one-loop_meg}
\end{align}
where the Yukawa couplings $y^e_{\phi\; ij}$ are defined in Eq.~(\ref{yukawa}).
Here we neglect an electron mass and we assume that the Yukawa coupling
$y^e_{\phi\; ee}$ is negligible.~\footnote{The Yukawa coupling $\rho_e^{ee}$ is strongly
constrained by $\tau \rightarrow \mu e^+ e^-$ process, as studied later. Therefore,
our assumption will be justified.}
The charged Higgs contribution to $A_{L,~R}$ is
\begin{align}
A_L^{H^-} &=-\frac{(\bar{\rho}_e^\dagger \rho_e)_{e\mu}}{192 \pi^2 m_{H^-}^2}, ~~A_R^{H^-}=0.
\end{align}
For nonzero $y^e_{\phi\; \mu e~(e\mu)}$, the Barr-Zee type contributions ($A_{L,R}^{\rm BZ}$) at two-loop level
are significant. The expression of $A_{L,~R}^{\rm BZ}$ is the same as one for $\tau \rightarrow \mu \gamma$ case
shown in Eq.~(\ref{Barr-Zee}) except that the flavor violating Yukawa couplings
$y^{e(*)}_{\phi\; \tau\mu~(\mu \tau)}$ should be replaced by $y^{e(*)}_{\phi\; \mu e ~(e\mu)}$,
and the $\tau$ mass ($m_\tau$) should be replaced by the $\mu$ mass ($m_\mu$).
The total $A_{L,~R}$ is a sum of all contributions;
\begin{align}
A_{L,~R}=\sum_{\phi=h,H,A,H^-} A_{L,~R}^\phi+A_{L,~R}^{\rm BZ}.
\end{align}

Similar to the muon g-2, the contributions from the $\mu-\tau$ flavor violating Yukawa interactions
together with the $e-\tau$ flavor violation have $O(m_\tau/m_\mu)$ enhancement, and induce significant
contributions to $\mu\rightarrow e\gamma$.
\begin{figure}
  \begin{center}
    %    {\epsfig{figure=FIGs/muonG2.pdf,width=0.6\textwidth}}
    \includegraphics[width=0.6\textwidth]{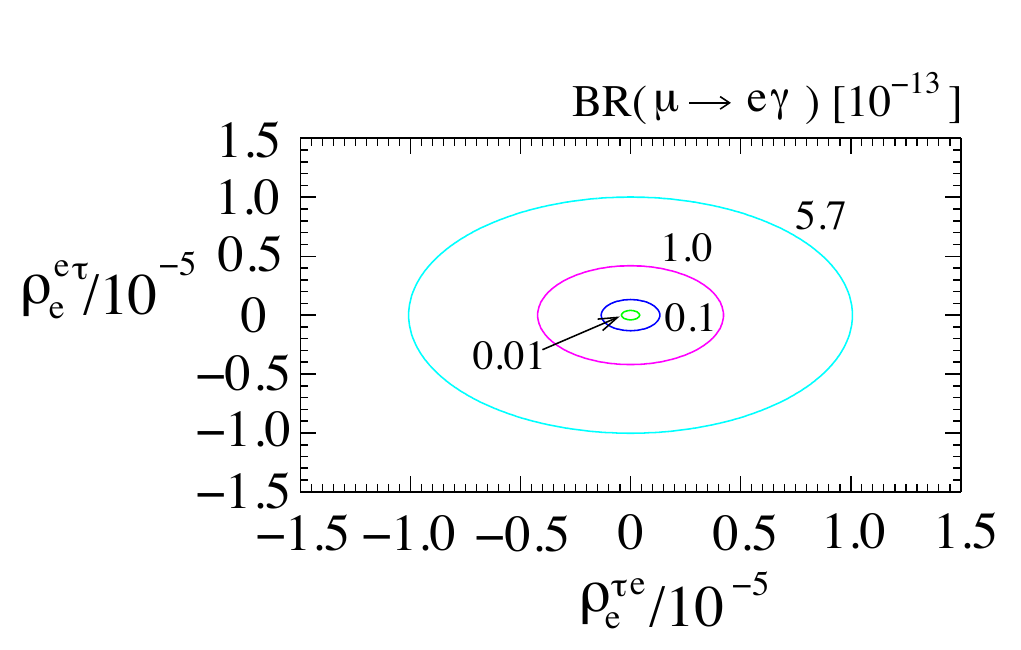}
    \caption{Numerical result for ${\rm BR}(\mu\rightarrow e\gamma)$ as a function of
      $\rho_e^{\tau e}$ and $\rho_e^{e\tau}$. Lines for ${\rm BR}(\mu \rightarrow e\gamma)/10^{-13}=5.7$
      (current limit), $1.0$, $0.1$ and $0.01$ are shown. Here we have assumed that $m_A=350$ GeV
      with $\lambda_4=\lambda_5=0.5$, $c_{\beta\alpha}=-0.007$ and
      ${\rm BR}(h\rightarrow \mu\tau)=0.84\%$ with $\rho_e^{\mu\tau}=-\rho_e^{\tau\mu}$, and
      extra Yukawa couplings $\rho_f$ other than $\rho_e^{\mu\tau(\tau\mu)}$ and $\rho_e^{\tau e(e\tau)}$
      are negligible.
      We note that for this parameter set, $\delta a_\mu=2.2\times 10^{-9}$.}
    \label{meg_tau-e}
  \end{center}
\end{figure}
In Fig.~\ref{meg_tau-e}, we show numerical results for ${\rm BR}(\mu\rightarrow e \gamma)$ as a function
of $\rho_e^{\tau e}$ and $\rho_e^{e\tau}$. Here we have taken $m_A=350$ GeV with $\lambda_4=\lambda_5=0.5$,
$c_{\beta\alpha}=-0.07$ and ${\rm BR}(h\rightarrow \mu\tau)=0.84~\%$ with $\rho_e^{\mu\tau}=-\rho_e^{\tau\mu}$.
We have assumed that extra Yukawa couplings $\rho_f^{ij}$ other than $\rho_e^{\mu\tau(\tau\mu)}$ and $\rho_e^{\tau e(e\tau)}$
are negligible. This parameter set corresponds to $\delta a_\mu=2.2\times 10^{-9}$. One can see that
the current limit on ${\rm BR}(\mu \rightarrow e\gamma)$ strongly constrains the $e-\tau$ flavor violating
couplings $\rho_e^{\tau e(e\tau)}$ if the CMS excess of ${\rm BR}(h \rightarrow \mu \tau)$ is true.
If we change the value of ${\rm BR}(h \rightarrow \mu\tau)$ in Fig.~\ref{meg_tau-e}, the experimental
bound of $\rho_e^{\tau e (e\tau)}$ is relaxed by the factor, $\sqrt{\frac{0.84\%}{{\rm BR}(h \rightarrow \mu\tau)}}$
when $\rho_e^{\tau e}=\rho_e^{e\tau}$ is assumed.

If Yukawa couplings $\rho_e^{\tau e (e\tau)}$ are negligible but  $\rho_e^{\mu e (e\mu)}$ are not,
the Barr-Zee type two loop contributions are dominant.\footnote{
  If $\rho_e^{\mu\mu}$ is also nonzero, there are also one-loop contributions as shown in Eq.~(\ref{one-loop_meg}).
  However, the coupling $\rho_e^{\mu\mu}$ is strongly constrained by the $\tau \rightarrow 3\mu$ bound, as discussed
  later. Therefore, the effect from $\rho_e^{\mu \mu}$ is negligible and we neglect it in our numerical analysis.}
We show numerical results
for ${\rm BR}(\mu \rightarrow e\gamma)$ as a function of $\rho_e^{\mu e}$ and $\rho_e^{e\mu}$.
\begin{figure}
  \begin{center}
    %    {\epsfig{figure=FIGs/muonG2.pdf,width=0.6\textwidth}}
    \includegraphics[width=0.6\textwidth]{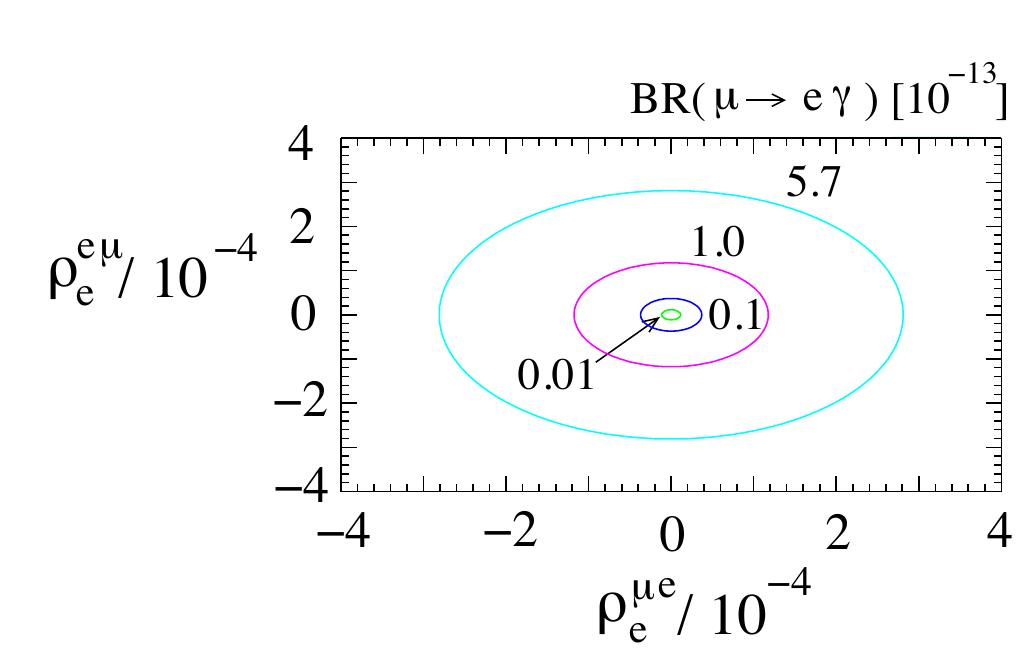}
    \caption{Numerical result for ${\rm BR}(\mu\rightarrow e\gamma)$ as a function of
      $\rho_e^{\mu e}$ and $\rho_e^{e\mu}$. Lines for ${\rm BR}(\mu \rightarrow e\gamma)/10^{-13}=5.7$
      (current limit), $1.0$, $0.1$ and $0.01$ are shown. Here we have assumed that $m_A=350$ GeV
      with $\lambda_4=\lambda_5=0.5$, $c_{\beta\alpha}=-0.007$,
      %      ${\rm BR}(h\rightarrow \mu\tau)=0.84\%$ with $\rho_e^{\mu\tau}=-\rho_e^{\tau\mu}$,
      and extra Yukawa couplings $\rho_f$ other than $\rho_e^{\mu\tau(\tau\mu)}$ and $\rho_e^{\mu e(e\mu)}$
      are negligible.}
    \label{meg_mu-e}
  \end{center}
\end{figure}
Here we have assumed that $m_A=350$ GeV with $\lambda_4=\lambda_5=0.5$, $c_{\beta \alpha}=-0.007$ and
extra Yukawa couplings $\rho_f^{ij}$ other than $\rho_e^{\mu\tau(\tau\mu)}$ and $\rho_e^{\mu e(e\mu)}$
are negligible. As one can see from the figure, $\rho_e^{\mu e(e\mu)}$ couplings are also severely 
constrained by the $\mu \rightarrow e\gamma$ bound. Note that the prediction of $\mu \rightarrow e \gamma$
for this case does not depend on the value of ${\rm BR}(h\rightarrow \mu\tau)$.
The future improvement of ${\rm BR}(\mu \rightarrow e \gamma)$ at the level of $10^{-14}$ as proposed
by the MEG II experiment~\cite{Baldini:2013ke} would significantly probe the flavor structure of this scenario.

The effective operator for $\mu \to e \gamma$ also generate $\mu-e$ conversion process in nuclei.
Besides, the extra Yukawa couplings, $\rho^{\mu e}_e$ and $\rho^{e \mu}_e$, may enhance the $\mu-e$ conversion
through the tree-level Higgs exchanging. The contribution depends on the extra Yukawa couplings in the quark sector as well, and then our model may be also tested by the experiments \cite{mu-e1, mu-e2,mu-e3,mu-e4},
although our prediction is vague because of the ambiguity of the Yukawa couplings.\footnote{The study on the tree-level flavor changing couplings of quarks is beyond our scope.}

We comment on the consequence of the strong constraints on the $e-\tau$ and $e-\mu$ flavor-violations.
Unlike the $\mu-\tau$ flavor violation, the $e-\tau$ flavor violating Yukawa couplings in this scenario is
strongly constrained as we have seen above. Therefore, the prediction of ${\rm BR}(\tau \rightarrow e \gamma)$
is expected to be small. Similarly, because of the smallness of the $e-\tau$ and $e-\mu$ flavor violation,
we also expect that the new physics contributions to the anomalous magnetic moment of electron (electron g-2)
should be small.

\subsection{Muon electric dipole moment (muon EDM)}
When we discussed the muon g-2, we have assumed that the $\mu-\tau$ flavor violating
Yukawa couplings are real. If the $\mu-\tau$ Yukawa couplings are complex, the couplings $\rho_e^{\mu\tau}\rho_e^{\tau\mu}$ in Eq.~(\ref{a_mu})
should be replaced by ${\rm Re}(\rho_e^{\mu\tau}\rho_e^{\tau\mu})$. In addition, the imaginary parts of the Yukawa couplings generate
an electric dipole moment (EDM) of muon. Since the muon g-2 and the muon EDM are induced by the same Feynman diagram shown in Fig.~\ref{muonG2},
these quantities are correlated via the unknown CP-violating phase.
The effective operators for the muon g-2 $(\delta a_\mu)$ and the muon EDM $(\delta d_\mu)$ are expressed by
\begin{align}
  {\cal L}=\bar{\mu}\sigma^{\mu\nu}\left( \frac{e}{4 m_\mu}\delta a_\mu-\frac{i}{2}\delta d_\mu\gamma_5\right)
  \mu F_{\mu \nu}.
\end{align}
If we parametrize the complex Yukawa couplings as follows:
\begin{align}
\rho_e^{\mu\tau}\rho_e^{\tau\mu}=|\rho_e^{\mu\tau}\rho_e^{\tau\mu}| e^{i\phi},
\end{align}
the relation between the muon g-2 ($\delta a_\mu$) and the muon EDM ($\delta d_\mu$) induced by the $\mu-\tau$ flavor-violating
Yukawa couplings is given by
\begin{align}
\frac{\delta d_\mu}{\delta a_\mu}=-\frac{e  \tan\phi}{2 m_\mu}.
\end{align}
Therefore, the predicted muon EDM is
\begin{eqnarray}
  \delta d_\mu =-3\times 10^{-22}~e\cdot {\rm cm}~\times \left(\frac{\tan \phi}{1.0}\right)
  \left(\frac{\delta a_\mu}{3\times 10^{-9}}\right).
\end{eqnarray}
The current limit~\cite{Bennett:2008dy} is
\begin{align}
|d_\mu|<1.9\times 10^{-19}~e\cdot {\rm cm}~(95\%~{\rm C.L.}),
\end{align}
and hence it is not sensitive to this scenario at present. However, the future improvement at the level
of $10^{-24}~e\cdot {\rm cm}$~\cite{Semertzidis:1999kv} would be significant to probe the scenario.

\subsection{$\tau \rightarrow \mu \nu \bar{\nu}$}
The Yukawa couplings $\rho_e^{\mu\tau(\tau\mu)}$ induce a correction
to $\tau \rightarrow \mu \nu\bar{\nu}$ via a charged Higgs mediation, where the flavor of final neutrino and anti-neutrino states is summed up since it is not detected.\footnote{In general, the unknown Yukawa couplings $\rho_e^{i \tau}$ and
  $\rho_e^{ i \mu}~(i=e,~\mu,~\tau)$ generate the extra corrections to $\delta$. However, the Yukawa couplings
  $\rho_e^{e\tau ~(e\mu)}$ and $\rho_e^{\mu\mu}$ are strongly constrained by $\mu \rightarrow e\gamma$ and
  $\tau^- \rightarrow \mu^- \mu^+\mu^-$, respectively. Therefore, the contributions from these couplings
  are negligible. The unknown Yukawa coupling $\rho_e^{\tau\tau}$ can be sizable, and hence it can increase
  the prediction of the $\delta$. Thus our result of $\delta$ induced from $\rho_e^{\mu\tau~(\tau\mu)}$
  is viewed as a conservative estimate.}
  
The correction $\delta$ is given as follows;
\begin{eqnarray}
  \Gamma(\tau \rightarrow \mu \nu \bar{\nu})&=&\frac{m_\tau^5 G_F^2}{192\pi^3}(1+\delta),\nonumber \\
  \delta &=&\frac{|\rho_e^{\mu\tau}|^2|\rho_e^{\tau\mu}|^2}{32G_F^2 m_{H^+}^4}.
  \label{taudecay_eq}
\end{eqnarray}
\begin{figure}
  \begin{center}
    %    {\epsfig{figure=note_FIG/delta_taudecay_case1.eps,width=0.8\textwidth}}
    \includegraphics[width=0.7\textwidth]{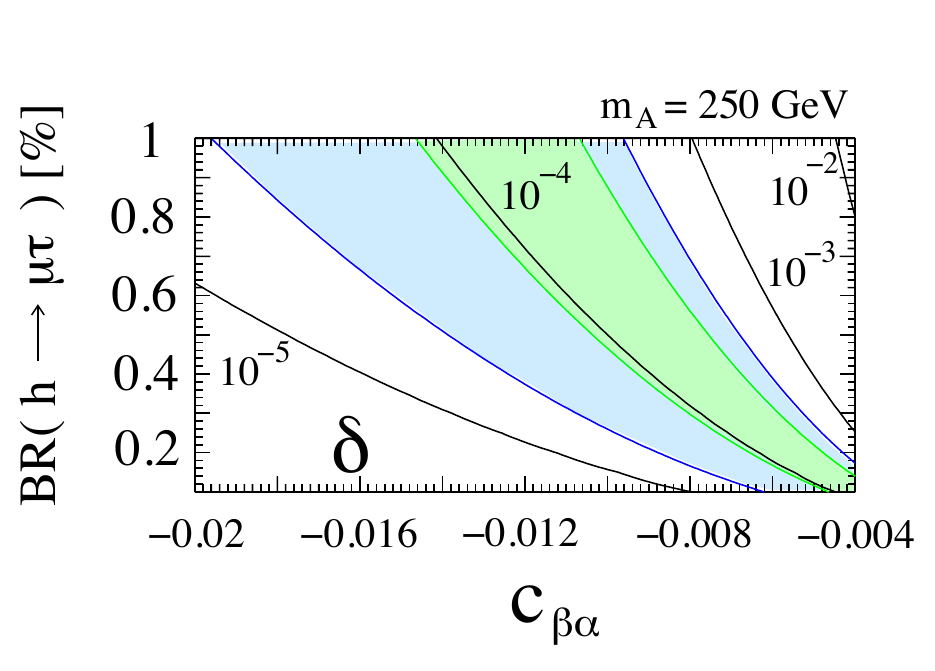}
    \includegraphics[width=0.7\textwidth]{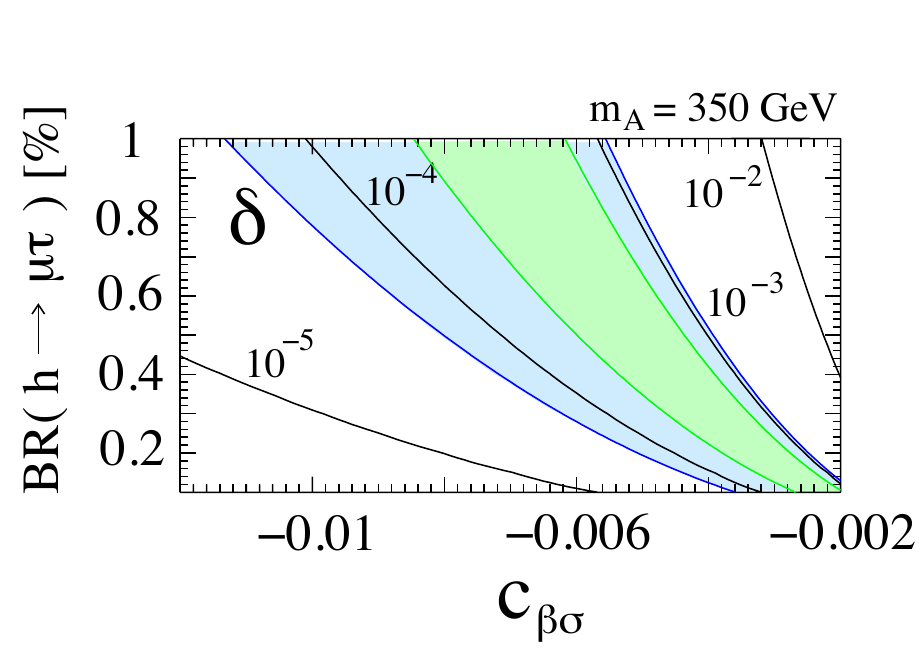}
    \caption{Numerical result for a correction to $\tau \rightarrow \mu \nu\bar{\nu}$,
      $\delta$ given in Eq.~(\ref{taudecay_eq}) as a function of $c_{\beta\alpha}$ and
      ${\rm BR}(h\rightarrow \mu\tau)$ in the same parameter set of
      Fig.~\ref{muonG2_vs_htm}.
      Black solid lines for $\delta=10^{-5},~10^{-4},~10^{-3},~10^{-2}$
      (from left to right) are shown for $m_A=250$ GeV (upper figure)
      and $m_A=350$ GeV (lower figure). Here we also show the region where
      the muon g-2 anomaly is explained within $1\sigma$ (light green) and
      $2\sigma$ (light blue).
      Here we have assumed that the extra Yukawa couplings $\rho_f$ other than
      $\rho_e^{\mu\tau(\tau\mu)}$ are negligible.}
    \label{taudecay}
  \end{center}
\end{figure}
In Fig.~\ref{taudecay}, numerical results for the correction $\delta$ given above
are shown as a function of $c_{\beta\alpha}$ and ${\rm BR}(h \rightarrow \mu\tau)$ in the same parameter
set of Fig.~\ref{muonG2_vs_htm}. One can see that as the correction to the muon g-2 ($\delta a_\mu$) gets
larger, the size of $\delta$ also becomes larger, and they are correlated each other, independent of
${\rm BR}(h\rightarrow \mu\tau)$. The interesting regions
which explain the muon g-2 anomaly within $1\sigma$ predict $\delta \le 10^{-4}$-$10^{-3}$. The current precision of
the measurement of the decay rate $\Gamma(\tau \rightarrow \mu\nu\bar{\nu})$ is at the level of
$10^{-3}$~\cite{Agashe:2014kda}.
Therefore, the further improvement of the precision would be important for this scenario. In addition,
from the $\tau$ decay, the BaBar collaboration has reported a measurement of the charged current lepton universality~\cite{Aubert:2009qj},
given by
\begin{align}
  \left(\frac{g_\mu}{g_e}\right)^2=\frac{{\rm BR}(\tau^-\rightarrow \mu^-\nu \bar{\nu})}{{\rm BR}
    (\tau^-\rightarrow e^- \nu \bar{\nu})}\frac{f(m_e^2/m_\tau^2)}{f(m_\mu^2/m_\tau^2)},
\label{lepton_univ}
\end{align}
where $f(x)=1-8x+8x^3-x^4-12x^2\log x$, which is a phase space factor. The universality of the gauge interaction in
the SM predicts $g_e=g_\mu$ and the current experimental results are
\begin{align}
  \left(\frac{g_\mu}{g_e}\right)&=1.0036\pm 0.0020~({\rm BaBar}), \nonumber \\
  &= 1.0018\pm 0.0014~({\rm world~ average}).
\end{align}
In our scenario, we expect the correction to $\tau \rightarrow e \nu \bar{\nu}$ would be
small because of the strong constraint on $e-\tau$ flavor violation
from $\mu \rightarrow e \gamma$ process. Therefore, the charged
Higgs contribution to $\tau \rightarrow \mu \nu \bar{\nu}$ with $\mu-\tau$ flavor violating Yukawa couplings
induces the significant correction to the violation of the lepton universality above,
\begin{align}
\left(\frac{g_\mu}{g_e}\right)^2=1+\delta.
\end{align}
The result from the Belle collaboration and the further improvement of the precision of the lepton
universality would have an important impact on our scenario.

\subsection{$\tau^-\rightarrow \mu^- l^+ l^-$, $\tau^-\rightarrow e^- l^+ l^-$
  $(l=e,~\mu)$, $\mu^+\rightarrow e^+ e^- e^+$ and others}
\begin{figure}
\begin{center}
  \includegraphics[width=0.7\textwidth]{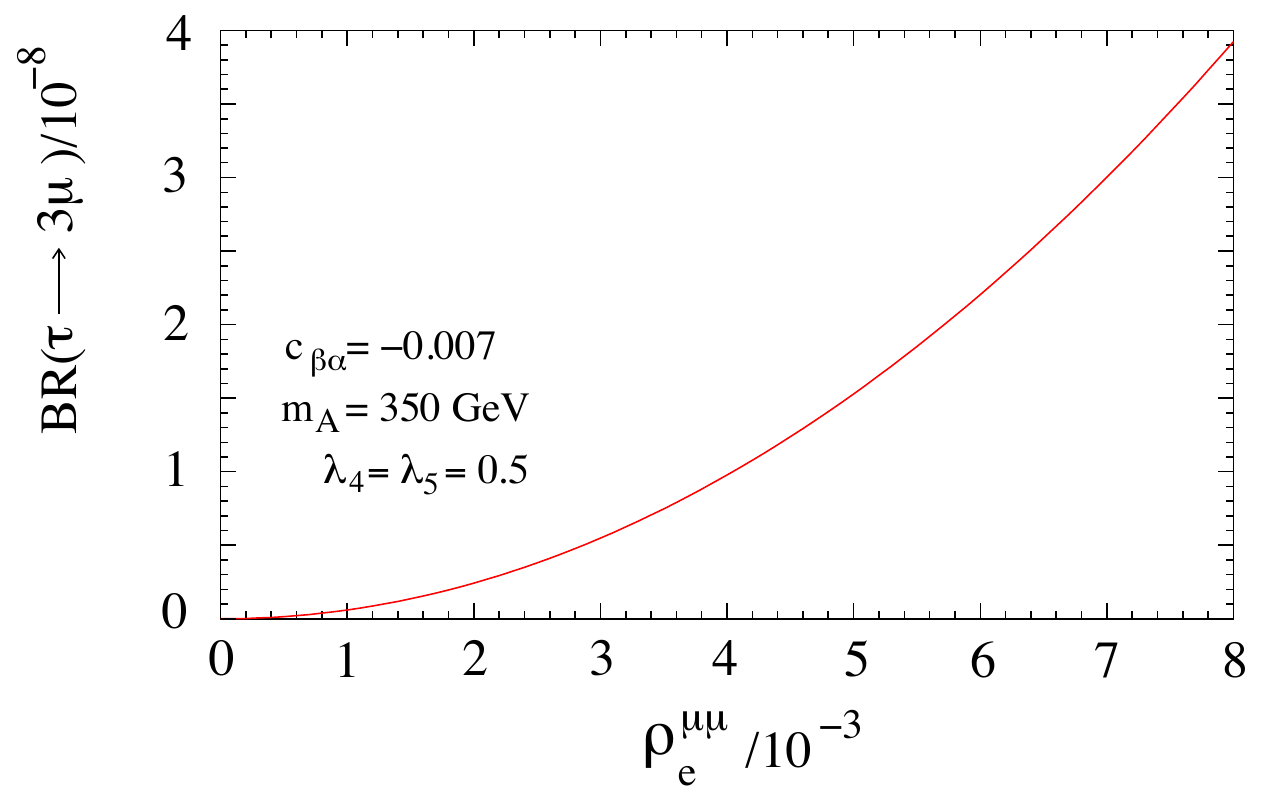}
  \includegraphics[width=0.7\textwidth]{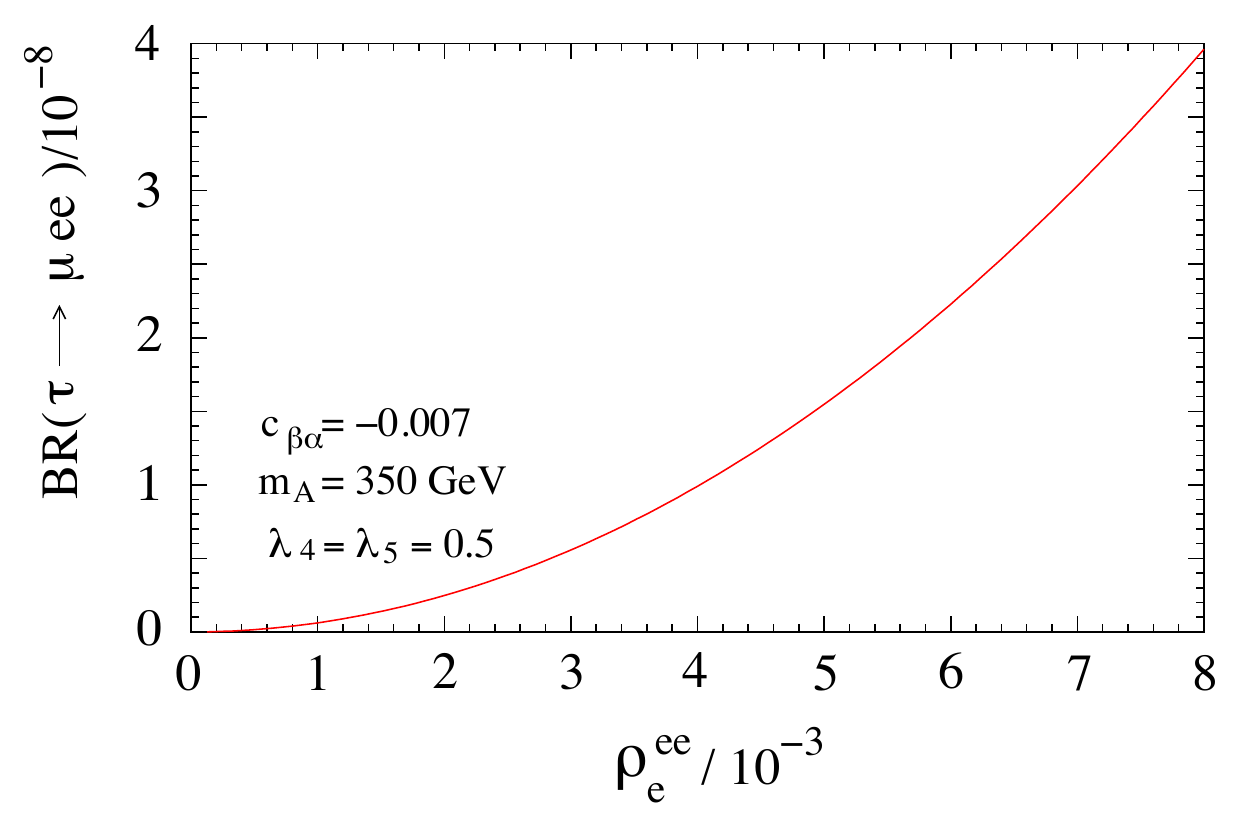}
  \caption{${\rm BR}(\tau\rightarrow 3\mu)$ (above) and ${\rm BR}(\tau \rightarrow \mu ee)$ (below)
    as a function of $\rho_e^{ll}$ ($l=\mu$ for $\tau \rightarrow 3\mu$ and $l=e$ for
    $\tau \rightarrow \mu ee$). Here we have assumed that $c_{\beta\alpha}=-0.007$, $m_A=350$ GeV
    with $\lambda_4=\lambda_5=0.5$ and
    ${\rm BR}(h\rightarrow \mu\tau)=0.84\%$ with $\rho_e^{\mu\tau}=-\rho_e^{\tau\mu}$.}
\label{tau_to_3mu}
\end{center}
\end{figure}

The nonzero Yukawa couplings $\rho_e^{\mu\tau(\tau\mu)}$ also generate processes $\tau^-\rightarrow \mu^-\mu^+\mu^-$
and $\tau^-\rightarrow \mu^- e^+ e^-$ ($ \tau \rightarrow 3\mu$ and $\tau \rightarrow \mu ee$ for short, respectively)
at the tree level.
They are induced without unknown $\rho_e^{\mu \mu}$ and $\rho_e^{ee}$ Yukawa couplings. The branching
ratios, however, are too small to be observed. Therefore, nonzero $\rho_e^{\mu\mu}$ and $\rho_e^{ee}$ are important for
these processes\footnote{
  Nonzero $\rho_e^{\tau e(e\tau)}$ and $\rho_e^{\mu e(e\mu)}$ couplings also induce the $\tau \rightarrow \mu ee$ process.
  However, these Yukawa couplings are strongly constrained by $\mu \rightarrow e \gamma$ process as discussed
  in previous sections. Therefore, we neglect these effects.}.
The branching ratios for $\tau \rightarrow 3\mu$ and $\tau\rightarrow \mu e e$ are given by~\cite{Crivellin:2013wna}
\begin{align}
\frac{{\rm BR}(\tau\rightarrow 3\mu)}{{\rm BR}(\tau \rightarrow \mu \nu\bar{\nu})}&=\sum_{\phi,~\phi'=h,~H,~A}\frac{I(\phi,~\phi')}{64G_F^2 },
\nonumber \\
I(\phi,~\phi')&=
2\left(\frac{y^e_{\phi \mu\tau} y^{e*}_{\phi\mu\mu}}{m_\phi^2}\right)\left(\frac{y^{e*}_{\phi' \mu\tau} y^e_{\phi' \mu \mu}}{m_{\phi'}^2}\right)
+2\left(\frac{y^{e}_{\phi\tau\mu} y^{e*}_{\phi\mu\mu}}{m_\phi^2}\right)\left(\frac{y^{e*}_{\phi' \tau \mu} y^{e}_{\phi'\mu\mu}}{m_{\phi'}^2}\right)
\nonumber \\
&+\left(\frac{y^e_{\phi \mu\tau} y^{e}_{\phi\mu\mu}}{m_\phi^2}\right)\left(\frac{y^{e*}_{\phi' \mu\tau} y^{e*}_{\phi' \mu \mu}}{m_{\phi'}^2}\right)
+\left(\frac{y^{e}_{\phi \tau\mu} y^{e}_{\phi\mu\mu}}{m_\phi^2}\right)\left(\frac{y^{e*}_{\phi'\tau \mu} y^{e*}_{\phi' \mu \mu}}{m_{\phi'}^2}\right),\\
\frac{{\rm BR}(\tau \rightarrow \mu ee)}{{\rm BR}(\tau \rightarrow \mu \nu \bar{\nu})}
&=\sum_{\phi,~\phi'=h,~H,~A}\frac{J(\phi,~\phi')}{32 G_F^2},
\nonumber \\
J(\phi,~\phi') &=
\left(\frac{y^e_{\phi \mu\tau} y^{e*}_{\phi ee}}{m_\phi^2}\right)\left(\frac{y^{e*}_{\phi' \mu\tau} y^e_{\phi' ee}}{m_{\phi'}^2}\right)
+\left(\frac{y^{e}_{\phi\tau\mu} y^{e*}_{\phi ee}}{m_\phi^2}\right)\left(\frac{y^{e*}_{\phi' \tau \mu} y^{e}_{\phi' ee}}{m_{\phi'}^2}\right)
\nonumber \\
&+\left(\frac{y^e_{\phi \mu\tau} y^{e}_{\phi ee}}{m_\phi^2}\right)\left(\frac{y^{e*}_{\phi' \mu\tau} y^{e*}_{\phi' ee}}{m_{\phi'}^2}\right)
+\left(\frac{y^{e}_{\phi \tau\mu} y^{e}_{\phi ee}}{m_\phi^2}\right)\left(\frac{y^{e*}_{\phi'\tau \mu} y^{e*}_{\phi' ee}}{m_{\phi'}^2}\right).
\end{align}
Fig.~\ref{tau_to_3mu} shows ${\rm BR}(\tau \rightarrow 3\mu)$ and ${\rm BR}(\tau \rightarrow \mu ee)$
as a function of $\rho_e^{ll}$ ($l=\mu$ for $\tau \rightarrow 3\mu$ and $l=e$ for $\tau \rightarrow \mu ee$).
It is assumed that $c_{\beta\alpha}=-0.007$, $m_A=350$ GeV
    with $\lambda_4=\lambda_5=0.5$ and
    ${\rm BR}(h\rightarrow \mu\tau)=0.84\%$ with $\rho_e^{\mu\tau}=-\rho_e^{\tau\mu}$ in Fig.~\ref{tau_to_3mu}.
One can see that the current experimental bounds,
\begin{eqnarray}
{\rm BR}(\tau\rightarrow 3\mu)<2.1\times 10^{-8},~~{\rm BR}(\tau \rightarrow \mu ee)<1.8\times 10^{-8}
\end{eqnarray}
set the strong constraints on the $\rho_e^{ll}$ Yukawa couplings. For example, the parameter set shown in
Fig.~\ref{tau_to_3mu}, requires $\rho_e^{ll}<0.006$~($l=\mu,~e$). We note that the constraint on the $\rho_e^{\mu\mu}$
is still larger than the value of the muon Yukawa coupling in the SM ($y_\mu=\frac{\sqrt{2}m_\mu}{v}\sim 6\times 10^{-4}$).

Contrary to $\tau^- \rightarrow \mu^- l^+ l^-$, the $\tau^- \rightarrow e^- l^+ l^-$ $(l=e,~\mu)$ process 
is suppressed in this scenario because the $\tau-e$ flavor violation is strongly constrained
by $\mu \rightarrow e \gamma$ process. Furthermore, since the constraints on $\rho_e^{e\mu~(\mu e)}$ are
stronger than those on $\rho_e^{\mu\mu(ee)}$, $\tau^-\rightarrow \mu^- e^+ \mu^-$ is expected to
be smaller than $\tau^- \rightarrow \mu^- l^+ l^-$. (Needless to say, $\tau^-\rightarrow e^- \mu^+ e^-$
is much suppressed.) Therefore, we expect that the $\tau^-\rightarrow e^- l^+ l^{-}$ and $\tau^- \rightarrow
\mu^- e^+ \mu^-$ processes will be small.

We also study $\mu^+ \rightarrow e^+ e^- e^+$ ($\mu \rightarrow 3e$ in short)
which depends on the $\mu-e$ flavor violating Yukawa couplings
$\rho_e^{e\mu~(\mu e)}$ and the flavor diagonal element $\rho_e^{ee}$.
As we have seen, the $\mu-e$ flavor violating Yukawa couplings $\rho_e^{e\mu (\mu e)}$ are
constrained by the $\mu \rightarrow e \gamma$ process and the $\rho_e^{ee}$ coupling is restricted by
the $\tau\rightarrow \mu ee$ process. From Fig.~\ref{meg_mu-e} and Fig.~\ref{tau_to_3mu}, the current limits on
$\rho_e^{\mu e (e \mu)}$ and $\rho_e^{ee}$ are
$\rho_e^{\mu e}<2\times 10^{-4}$ for $\rho_e^{\mu e}=\rho_e^{e\mu}$ and $\rho_e^{e e}<6\times 10^{-3}$, respectively,
assuming $m_A=350$ GeV with $\lambda_4=\lambda_5=0.5$ and $c_{\beta\alpha}=-0.007$.
Under these constraints, it will be interesting to see how large branching ratio of $\mu \rightarrow 3e$
is expected.
\begin{figure}
\begin{center}
  \includegraphics[width=0.7\textwidth]{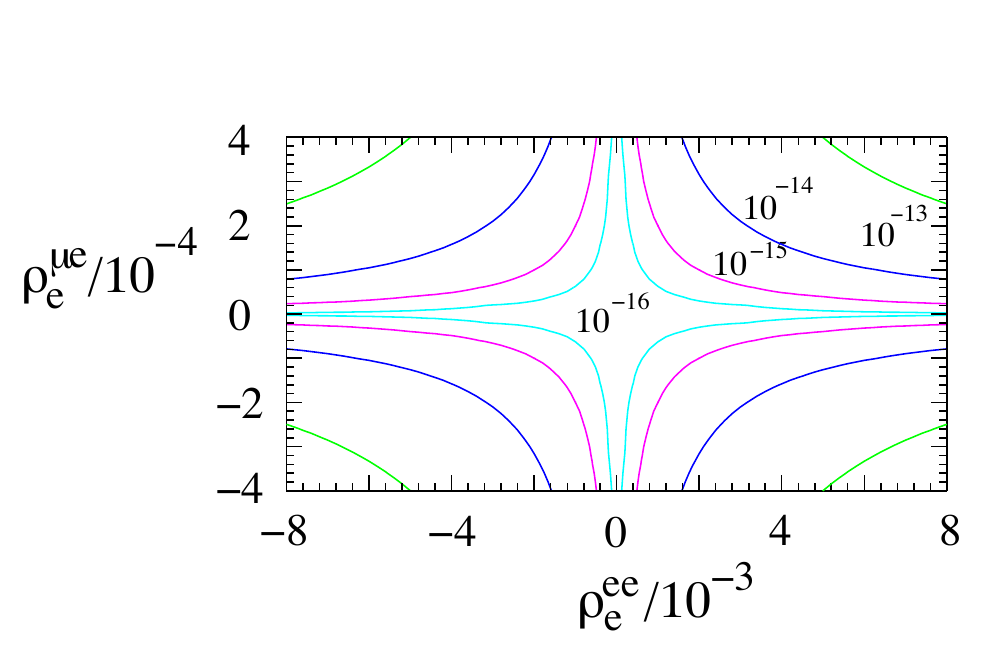}
  \caption{${\rm BR}(\mu\rightarrow 3 e)$ 
    as a function of $\rho_e^{ee}$ and $\rho_e^{\mu e}$. Here we have assumed that $\rho_e^{\mu e}=\rho_e^{e\mu}$,
    $c_{\beta\alpha}=-0.007$ and $m_A=350$ GeV with $\lambda_4=\lambda_5=0.5$.}
\label{mu_to_3e}
\end{center}
\end{figure}
In Fig.~\ref{mu_to_3e}, we show the ${\rm BR}(\mu \rightarrow 3e)$ as a function of $\rho_e^{ee}$ and
$\rho_e^{\mu e}$. In the parameter region where the constraints from $\mu \rightarrow e\gamma$ and
$\tau\rightarrow \mu ee$ are satisfied, the branching ratio can be as large as about $10^{-13}$.
This is consistent with the current limit~\cite{Agashe:2014kda}
\begin{align}
{\rm BR}(\mu \rightarrow 3e)<1.0\times 10^{-12}.
\end{align}
The improvement of the branching ratio at the level of $10^{-16}$~\cite{Mu3e}
which has been proposed by the Mu3e experiment would have a significant impact on this scenario together with the improvement of
$\mu \rightarrow e \gamma$~\cite{Baldini:2013ke} and $\mu -e$ conversion in nuclei \cite{mu-e1, mu-e2,mu-e3,mu-e4}.

\subsection{$\tau\rightarrow \mu\eta$}
The $\tau \rightarrow \mu \eta$ is also generated by the extra $\rho_d^{ss}$ Yukawa coupling via the mediation of
the CP-odd Higgs boson at the tree level.
The expression for the branching ratio of $\tau \rightarrow \mu\eta$ is given by~\cite{Sher:2002ew,Black:2002wh}
\begin{align}
  {\rm BR}(\tau\rightarrow \mu\eta)=\frac{3|\rho_d^{ss}|^2 (\bar{\rho}^{\mu\tau})^2}{32\pi}
  \frac{m_\tau F_\eta^2}{m_A^4 \Gamma_\tau}\left(
  \frac{m_\eta^2}{m_u+m_d+4m_s}\right)^2 \left(1-\frac{m_\eta^2}{m_\tau^2}\right)^2,
\end{align}
where $m_\eta$ and $F_\eta$ are the mass and the decay constant of $\eta$.
For $F_\eta=150$ MeV and $m_\eta=548$ MeV, we obtain a constraint on $\rho_d^{ss}$;
\begin{eqnarray}
  |\rho_d^{ss}|<0.007 \left(\frac{0.3}{\bar{\rho}^{\mu\tau}}\right)
  \left(\frac{m_A}{350~{\rm GeV}}\right)^2.
\end{eqnarray}
We have a strong constraint although it is still larger than
the SM value of the strange quark Yukawa coupling ($y_s=\frac{\sqrt{2}m_s}{v}\sim 5\times 10^{-4}$).

The other hadronic $\tau$-lepton decays have been studied in Ref.~\cite{Celis:2013xja}.
They potentially
provide constraints on the other extra Yukawa couplings $\rho_f$ in quark sector. For details,
see Ref.~\cite{Celis:2013xja}.

\section{Implication to Higgs physics}
We have seen that the CMS excess in $h\rightarrow \mu \tau$ is consistent with
the anomaly of muon g-2 as well as the other experimental constraints. It will be interesting to
note whether other lepton flavor violating Higgs boson decays would be possible.
As we have already seen, the $e-\mu$ and $e-\tau$ flavor violating Yukawa couplings are strongly
constrained mainly by the $\mu \rightarrow e \gamma$ constraint. As a consequence, the lepton flavor
violating Higgs boson decays $h\rightarrow e\mu$ and $h\rightarrow e\tau$ are strongly suppressed
so that the near future experiments such as the ones at the LHC could not observe these decay modes,
contrary to the $h \rightarrow \mu \tau$ mode. Therefore, the non-observation of these decays
is one of interesting predictions of this scenario.

\section{Summary}
The anomalous event in $h \rightarrow \mu \tau$ has been observed by the CMS collaboration. 
The discrepancy of the muon g-2 is also one of the longstanding issues in the particle physics. 
These anomalous phenomena may be a hint of physics beyond the Standard Model.
 At glance, these anomalies are not related each others.
However, we have found that the both anomalies are related and accommodated by the $\mu-\tau$ flavor
violating Yukawa interactions in a general two Higgs doublet model, and hence this motivates
further studies to see whether there are any interesting predictions and indications in the scenario. 
We have identified the parameter space where the CMS excess in $h\rightarrow \mu\tau$ and
the muon g-2 anomaly are both explained, and especially we have studied $\tau$- and $\mu$- physics in this
interesting parameter space.

One of the interesting processes in the presence of the $\mu-\tau$ flavor violation is
$\tau \rightarrow \mu \gamma$. The $\mu-\tau$ flavor violation suggested by the CMS excess
in $h\rightarrow \mu\tau$ and the muon g-2 anomaly induces the large branching ratio, and it can be
as large as $10^{-9}$ which is within the reach of the future experiment at the SuperKEKB. The imaginary parts of
the $\mu-\tau$ flavor violating Yukawa couplings also induce the extra contributions to the muon EDM,
which may be also within the planning future experiments. The necessary $\mu-\tau$ flavor violation also
generates the correction to $\tau \rightarrow \mu \nu\bar{\nu}$ decay and also induces a violation of
lepton universality between $\tau\rightarrow \mu \nu\bar{\nu}$ and $\tau \rightarrow e\nu\bar{\nu}$.
The improvement of their precisions would be interesting. The tree-level $\tau$ decays such as
$\tau^- \rightarrow \mu^- l^+ l^-~(l=e,~\mu)$ and $\tau \rightarrow \mu \eta$ are also interesting because
the extra Yukawa couplings $\rho_e^{ee~(\mu\mu)}$ and $\rho_d^{ss}$ could also induce the observable effects.
On the other hand, we have found that the $e-\mu$ and
$e-\tau$ flavor violating Yukawa couplings are severely constrained by mainly $\mu \rightarrow e\gamma$
process. Because of these constraints, phenomena such as $\tau \rightarrow e\gamma$,
$\tau^- \rightarrow e^- l^+ l^{-}~(l,=e,~\mu),~e^-\mu^+e^-,~\mu^-e^+\mu^-$ and extra contributions
to the electron g-2 would not be accessible in the near future experiments. Although there are many unknown Yukawa couplings in a general 2HDM, there are many interesting indications to $\tau$- and $\mu$-physics.

We have also commented on an implication to Higgs physics. Contrary to the $\mu-\tau$ flavor violation
suggested by the CMS result, the $e-\mu$ and $e-\tau$ flavor violations in the Higgs coupling are strongly
limited. Therefore, the observation of $h\rightarrow \mu \tau$ and non-observation of $h\rightarrow e\mu$
and $h\rightarrow e\tau$ would be the important implication of the scenario. 

We summarize our findings in Table ~\ref{summary_table}. If the CMS excess in $h\rightarrow \mu \tau$
is justified in coming LHC run, these phenomena in $\tau$- and $\mu$- physics would be key to reveal
the physics beyond the Standard Model.

\acknowledgments
%\vspace{1cm}

This work was supported in part by Grants-in-Aid for Scientific Research from the Ministry of Education,
Culture, Sports, Science, and Technology (MEXT), Japan (No.22224003 for K.T. and No. 23104011 for Y.O.)
and Japan Society for Promotion of Science (JSPS) (No.26104705 for K.T.).
E.S. is supported in part by the Ministry of Science and Technology, Taiwan 
under Grant No. MOST 104-2811-M-008-011.

\begin{table}
\begin{center}
\begin{tabular}{|c|c|c|}
  \hline
  Process & typical value &  observability\\
  \hline
  muon g-2 & $\delta a_\mu=(2.6\pm 0.8)\times 10^{-9}$ & (input)\\
  $\tau \rightarrow \mu \gamma$ & ${\rm BR} \le 10^{-9}$  & $\circ$\\
  $\tau \rightarrow e \gamma$ & small & $\times$\\
  $\tau \rightarrow \mu l^+l^-~(l=e,~\mu)$ & depends on $\rho_e^{\mu\mu}$ and $\rho_e^{ee}$ & ($\circ$)\\
  $\tau^- \rightarrow e^- l^+ l^-,~e^- \mu^+ e^-,~\mu^- e^+ \mu^-$ & small & $\times$\\
  $\tau \rightarrow \mu \eta$ & depends on $\rho_d^{ss}$ & ($\circ$)\\
  $\tau \rightarrow \mu \nu \bar{\nu}$ & $\delta \le 10^{-3}$,~lepton non-universality & $\triangle$\\
  $\tau \rightarrow e \nu \bar{\nu}$ & small, lepton non-universality & $\triangle$\\
  $\mu \rightarrow e\gamma$ & depends on $\rho_e^{\tau e (e\tau)}$ and $\rho_e^{\mu e(e\mu)}$ & ($\circ$)\\
   $\mu-e$ conversion & depends on $\rho_e^{\mu e(e\mu)}$ and  $\rho_{d,u}^{ij}$ & ($\circ$)\\
  $\mu \rightarrow 3e$ & ${\rm BR} \le 10^{-13}$ & $(\circ)$\\
  muon EDM & $|\delta d_\mu| \le 10^{-22} e\cdot {\rm cm}$ & $(\circ)$\\
  electron g-2 & small & $\times$\\
  \hline\hline
  LFV Higgs decay mode & {\rm BR} & \\
  \hline
  $h\rightarrow \mu \tau$ & ${\rm BR}=(0.84^{+0.39}_{-0.37})$\% & (input) \\
  $h\rightarrow e \tau$ & small & $\times$\\
  $h\rightarrow e \mu$ & small & $\times$\\
  \hline
\end{tabular}
\end{center}
\caption{Observalilities in various processes are summarized.
  If there is an observability in the planning experiments without introducing unknown Yukawa couplings $\rho_f$
  other than $\rho_e^{\mu\tau~(\tau\mu)}$, the circle mark ``$\circ$'' is shown. If there is an observability, but it depends
  on unknown Yukawa couplings (other than $\rho_e^{\mu\tau~(\tau\mu)}$), ``$(\circ$)'' is indicated.
  If there is an observability when the (currently unknown) experimental improvement is achieved, the triangle
  mark ``$\triangle$'' is shown. If the event rate is expected to be too small to be observed, ``$\times$'' is
  shown.}
\label{summary_table}

\end{table}


\begin{thebibliography}{xx}
  %%\cite{Aad:2012tfa}
\bibitem{Aad:2012tfa} 
  G.~Aad {\it et al.} [ATLAS Collaboration],
``Observation of a new particle in the search for the Standard Model Higgs boson with the ATLAS detector at the LHC,''
  Phys.\ Lett.\ B {\bf 716}, 1 (2013)
  [arXiv:1207.7214 [hep-ex]].
  %%CITATION = ARXIV:1207.7214;%%
  %5016 citations counted in INSPIRE as of 08 Oct 2015
%\cite{Chatrchyan:2012xdj}
\bibitem{Chatrchyan:2012xdj} 
  S.~Chatrchyan {\it et al.} [CMS Collaboration],
  ``Observation of a new boson at a mass of 125 GeV with the CMS experiment at the LHC,''
  Phys.\ Lett.\ B {\bf 716}, 30 (2013)
  [arXiv:1207.7235 [hep-ex]].
  %%CITATION = ARXIV:1207.7235;%%
  %4921 citations counted in INSPIRE as of 08 Oct 2015  
  %\cite{Khachatryan:2015kon}
%\cite{Bjorken:1977vt}
\bibitem{Bjorken:1977vt} 
  J.~D.~Bjorken and S.~Weinberg,
``A Mechanism for Nonconservation of Muon Number,''
  Phys.\ Rev.\ Lett.\  {\bf 38}, 622 (1977).
  %%CITATION = PRLTA,38,622;%%
  %144 citations counted in INSPIRE as of 08 Oct 2015
%\cite{Glashow:1976nt}
\bibitem{Glashow:1976nt} 
  S.~L.~Glashow and S.~Weinberg,
  ``Natural Conservation Laws for Neutral Currents,''
  Phys.\ Rev.\ D {\bf 15}, 1958 (1977).
  %%CITATION = PHRVA,D15,1958;%%
  %1375 citations counted in INSPIRE as of 08 Oct 2015
%\cite{McWilliams:1980kj}
\bibitem{McWilliams:1980kj} 
  B.~McWilliams and L.~F.~Li,
``Virtual Effects of Higgs Particles,''
  Nucl.\ Phys.\ B {\bf 179}, 62 (1981).
  %%CITATION = NUPHA,B179,62;%%
  %110 citations counted in INSPIRE as of 08 Oct 2015
%\cite{Shanker:1981mj}
\bibitem{Shanker:1981mj} 
  O.~U.~Shanker,
 ``Flavor Violation, Scalar Particles and Leptoquarks,''
  Nucl.\ Phys.\ B {\bf 206}, 253 (1982).
  %%CITATION = NUPHA,B206,253;%%
  %131 citations counted in INSPIRE as of 08 Oct 2015
%\cite{Cheng:1987rs}
\bibitem{Cheng:1987rs} 
  T.~P.~Cheng and M.~Sher,
``Mass Matrix Ansatz and Flavor Nonconservation in Models with Multiple Higgs Doublets,''
  Phys.\ Rev.\ D {\bf 35}, 3484 (1987).
  %%CITATION = PHRVA,D35,3484;%%
  %458 citations counted in INSPIRE as of 08 Oct 2015
  
\bibitem{Khachatryan:2015kon}
  V.~Khachatryan {\it et al.} [CMS Collaboration],
  ``Search for lepton-flavour-violating decays of the Higgs boson,''
  arXiv:1502.07400 [hep-ex].
  %%CITATION = ARXIV:1502.07400;%%
  %28 citations counted in INSPIRE as of 29 juil. 2015
\bibitem{ATLAS_hmt}
  Talk by Pierre Savard in European Physical Society Conference on High Energy Physics 2015,
  27 July 2015, Vienna, Austria;
  The ATLAS Collaboration, ``Search for lepton-flavour-violating $H\rightarrow \mu\tau$
  decays of the Higgs boson with the ATLAS detector'', arXiv:1508.03372 [hep-ex].

\bibitem{Sierra:2014nqa} 
  D.~Aristizabal Sierra and A.~Vicente,
 ``Explaining the CMS Higgs flavor violating decay excess,''
  Phys.\ Rev.\ D {\bf 90}, no. 11, 115004 (2014)
  [arXiv:1409.7690 [hep-ph]].
  %%CITATION = ARXIV:1409.7690;%%
  %30 citations counted in INSPIRE as of 16 Oct 2015
  \bibitem{Heeck:2014qea} 
  J.~Heeck, M.~Holthausen, W.~Rodejohann and Y.~Shimizu,
``$H\rightarrow \mu\tau$ in Abelian and non-Abelian flavor symmetry models,''
  Nucl.\ Phys.\ B {\bf 896}, 281 (2015)
  [arXiv:1412.3671 [hep-ph]].

\bibitem{Crivellin:2015mga} 
  A.~Crivellin, G.~D’Ambrosio and J.~Heeck,
``Explaining $h\to\mu^\pm\tau^\mp$, $B\to K^* \mu^+\mu^-$ and $B\to K \mu^+\mu^-/B\to K e^+e^-$ in a two-Higgs-doublet model with gauged $L_\mu-L_\tau$,''
  Phys.\ Rev.\ Lett.\  {\bf 114}, 151801 (2015)
  [arXiv:1501.00993 [hep-ph]].

\bibitem{deLima:2015pqa} 
  L.~de Lima, C.~S.~Machado, R.~D.~Matheus and L.~A.~F.~do Prado,
``Higgs Flavor Violation as a Signal to Discriminate Models,''
  arXiv:1501.06923 [hep-ph].

\bibitem{Omura:2015nja} 
  Y.~Omura, E.~Senaha and K.~Tobe,
``Lepton-flavor-violating Higgs decay $h \to \mu\tau$ and muon anomalous magnetic moment in a general two Higgs doublet model,''
  JHEP {\bf 1505}, 028 (2015)
  [arXiv:1502.07824 [hep-ph]].

\bibitem{Das:2015kea} 
  S.~P.~Das, J.~Hern\'andez-S\'anchez, A.~Rosado and R.~Xoxocotzi,
``Flavor violating signatures of lighter and heavier Higgs bosons within Two Higgs Doublet Model type III at the LHeC,''
  arXiv:1503.01464 [hep-ph].

\bibitem{Crivellin:2015lwa} 
  A.~Crivellin, G.~D’Ambrosio and J.~Heeck,
``Addressing the LHC flavor anomalies with horizontal gauge symmetries,''
  Phys.\ Rev.\ D {\bf 91}, no. 7, 075006 (2015)
  [arXiv:1503.03477 [hep-ph]].
\bibitem{Das:2015zwa} 
  D.~Das and A.~Kundu,
``Two hidden scalars around 125 GeV and h→μτ,''
  Phys.\ Rev.\ D {\bf 92}, no. 1, 015009 (2015)
  [arXiv:1504.01125 [hep-ph]].

\bibitem{Yue:2015dia} 
  C.~X.~Yue, C.~Pang and Y.~C.~Guo,
``Lepton flavor violating Higgs couplings and single production of the Higgs boson via e$\gamma$ collision,''
  J.\ Phys.\ G {\bf 42}, 075003 (2015)
  [arXiv:1505.02209 [hep-ph]].

\bibitem{Bhattacherjee:2015sia} 
  B.~Bhattacherjee, S.~Chakraborty and S.~Mukherjee,
``$H \rightarrow \tau \mu$ and excess in $t\bar{t}H$: Connecting the dots in the hope for the first glimpse of BSM Higgs signal,''
  arXiv:1505.02688 [hep-ph].

\bibitem{Mao:2015hwa} 
  Y.~n.~Mao and S.~h.~Zhu,
``On the Higgs-$\mu$-$\tau$ Coupling at High and Low Energy Colliders,''
  arXiv:1505.07668 [hep-ph].

\bibitem{He:2015rqa} 
  X.~G.~He, J.~Tandean and Y.~J.~Zheng,
``Higgs decay $h\rightarrow \mu\tau$ with minimal flavor violation,''
  JHEP {\bf 1509}, 093 (2015)
  [arXiv:1507.02673 [hep-ph]].

\bibitem{Goto:2015iha} 
  T.~Goto, R.~Kitano and S.~Mori,
``Lepton flavor violating $Z$-boson couplings from non-standard Higgs interactions,''
  arXiv:1507.03234 [hep-ph].

\bibitem{Chiang:2015cba} 
  C.~W.~Chiang, H.~Fukuda, M.~Takeuchi and T.~T.~Yanagida,
``Flavor-Changing Neutral-Current Decays in Top-Specific Variant Axion Model,''
  arXiv:1507.04354 [hep-ph].

\bibitem{Crivellin:2015hha} 
  A.~Crivellin, J.~Heeck and P.~Stoffer,
``A perturbed lepton-specific two-Higgs-doublet model facing experimental hints for physics beyond the Standard Model,''
  arXiv:1507.07567 [hep-ph].

\bibitem{Cheung:2015yga} 
  K.~Cheung, W.~Y.~Keung and P.~Y.~Tseng,
  ``Leptoquark induced rare decay amplitudes $h \to \tau^\mp \mu^\pm$ and $\tau\to \mu\gamma$,''
  arXiv:1508.01897 [hep-ph].

\bibitem{Botella:2015hoa} 
  F.~J.~Botella, G.~C.~Branco, M.~Nebot and M.~N.~Rebelo,
  ``Flavour Changing Higgs Couplings in a Class of Two Higgs Doublet Models,''
  arXiv:1508.05101 [hep-ph].

\bibitem{Liu:2015oaa} 
  X.~Liu, L.~Bian, X.~Q.~Li and J.~Shu,
``$h\rightarrow\mu\tau$, muon g$-$2, and a possible interpretation of the Galactic Center gamma ray excess,''
  arXiv:1508.05716 [hep-ph].

\bibitem{Baek:2015mea} 
  S.~Baek and K.~Nishiwaki,
  ``Leptoquark explanation of $h \to \mu\tau$ and muon $(g-2)$,''
  arXiv:1509.07410 [hep-ph].

\bibitem{Huang:2015vpt} 
  W.~Huang and Y.~L.~Tang,
  ``Flavor anomalies at the LHC and the R-parity violating supersymmetric model extended with vectorlike particles,''
  Phys.\ Rev.\ D {\bf 92}, no. 9, 094015 (2015)
  [arXiv:1509.08599 [hep-ph]].

\bibitem{Baek:2015fma} 
  S.~Baek and Z.~F.~Kang,
  ``Naturally Large Radiative Lepton Flavor Violating Higgs Decay Mediated by Lepton-flavored Dark Matter,''
  arXiv:1510.00100 [hep-ph].
\bibitem{Arganda:2015uca} 
  E.~Arganda, M.~J.~Herrero, R.~Morales and A.~Szynkman,
  ``Analysis of the $h, H, A \to \tau \mu$ decays induced from SUSY loops within the Mass Insertion Approximation,''
  arXiv:1510.04685 [hep-ph].
  %%CITATION = ARXIV:1510.04685;%%
  %1 citations counted in INSPIRE as of 06 Nov 2015
  
     %%%%%%%%%%%%%%%%%%%%%%%%%%%
\bibitem{Aloni:2015wvn} 
  D.~Aloni, Y.~Nir and E.~Stamou,
  ``Large BR($h \to \tau \mu$) in the MSSM,''
  arXiv:1511.00979 [hep-ph].
  %%CITATION = ARXIV:1511.00979;%%
  
  
  
  
  
  
  
  
  

  %%%%%%%%%%%%%%%%%%%%%%%%%%%
  
%\cite{Assamagan:2002kf}
\bibitem{Assamagan:2002kf} 
  K.~A.~Assamagan, A.~Deandrea and P.~A.~Delsart,
``Search for the lepton flavor violating decay $A_0/H_0\rightarrow \tau^\pm \mu^\mp$ at hadron colliders,''
  Phys.\ Rev.\ D {\bf 67}, 035001 (2003)
  [hep-ph/0207302].
  %%CITATION = HEP-PH/0207302;%%
  %43 citations counted in INSPIRE as of 16 Oct 2015
%\cite{Brignole:2003iv}
\bibitem{Brignole:2003iv} 
  A.~Brignole and A.~Rossi,
``Lepton flavor violating decays of supersymmetric Higgs bosons,''
  Phys.\ Lett.\ B {\bf 566}, 217 (2003)
  [hep-ph/0304081].
  %%CITATION = HEP-PH/0304081;%%
  %129 citations counted in INSPIRE as of 16 Oct 2015
%\cite{Kanemura:2004cn}
\bibitem{Kanemura:2004cn} 
  S.~Kanemura, K.~Matsuda, T.~Ota, T.~Shindou, E.~Takasugi and K.~Tsumura,
``Search for lepton flavor violation in the Higgs boson decay at a linear collider,''
  Phys.\ Lett.\ B {\bf 599}, 83 (2004)
  [hep-ph/0406316].
  %%CITATION = HEP-PH/0406316;%%
  %40 citations counted in INSPIRE as of 16 Oct 2015
%\cite{Arganda:2004bz}
\bibitem{Arganda:2004bz} 
  E.~Arganda, A.~M.~Curiel, M.~J.~Herrero and D.~Temes,
``Lepton flavor violating Higgs boson decays from massive seesaw neutrinos,''
  Phys.\ Rev.\ D {\bf 71}, 035011 (2005)
  [hep-ph/0407302].
  %%CITATION = HEP-PH/0407302;%%
  %70 citations counted in INSPIRE as of 16 Oct 2015
%\cite{Kanemura:2005hr}
\bibitem{Kanemura:2005hr} 
  S.~Kanemura, T.~Ota and K.~Tsumura,
``Lepton flavor violation in Higgs boson decays under the rare tau decay results,''
  Phys.\ Rev.\ D {\bf 73}, 016006 (2006)
  [hep-ph/0505191].
  %%CITATION = HEP-PH/0505191;%%
  %61 citations counted in INSPIRE as of 16 Oct 2015
%\cite{Blankenburg:2012ex}
\bibitem{Blankenburg:2012ex} 
  G.~Blankenburg, J.~Ellis and G.~Isidori,
``Flavour-Changing Decays of a 125 GeV Higgs-like Particle,''
  Phys.\ Lett.\ B {\bf 712}, 386 (2014)
  [arXiv:1202.5704 [hep-ph]].
  %%CITATION = ARXIV:1202.5704;%%
  %86 citations counted in INSPIRE as of 16 Oct 2015  
  %\cite{Harnik:2012pb}
\bibitem{Harnik:2012pb} 
  R.~Harnik, J.~Kopp and J.~Zupan,
``Flavor Violating Higgs Decays,''
  JHEP {\bf 1303}, 026 (2013)
  [arXiv:1209.1397 [hep-ph]].
  %%CITATION = ARXIV:1209.1397;%%
  %102 citations counted in INSPIRE as of 11 sept. 2015
  

%\cite{Agashe:2014kda}
\bibitem{Agashe:2014kda}
  K.~A.~Olive {\it et al.} [Particle Data Group Collaboration],
  ``Review of Particle Physics,''
  Chin.\ Phys.\ C {\bf 38} (2014) 090001.
  %%CITATION = CHPHD,C38,090001;%%
  %1635 citations counted in INSPIRE as of 11 Aug 2015

  
  %%%%%%%%%%%%%%%%%%%%%%%
  
\bibitem{peskin}
%\cite{Peskin:1991sw}
%\bibitem{Peskin:1991sw} 
  M.~E.~Peskin and T.~Takeuchi,
  ``Estimation of oblique electroweak corrections,''
  Phys.\ Rev.\ D {\bf 46}, 381 (1992).
  %%CITATION = PHRVA,D46,381;%%
  %1288 citations counted in INSPIRE as of 03 Jul 2013
  
  
  %%%%%%%%%%%%%%%%%%%%%%
  
  
  
  
  
%\cite{Hagiwara:2011af}
\bibitem{Hagiwara:2011af} 
  K.~Hagiwara, R.~Liao, A.~D.~Martin, D.~Nomura and T.~Teubner,
``$(g-2)_\mu$ and $\alpha(M_Z^2)$ re-evaluated using new precise data,''
  J.\ Phys.\ G {\bf 38}, 085003 (2011)
  [arXiv:1105.3149 [hep-ph]].
  %%CITATION = ARXIV:1105.3149;%%
  %345 citations counted in INSPIRE as of 16 Oct 2015
%\cite{Kanemitsu:2012dc}
\bibitem{Kanemitsu:2012dc} 
  S.~Kanemitsu and K.~Tobe,
``New physics for muon anomalous magnetic moment and its electroweak precision analysis,''
  Phys.\ Rev.\ D {\bf 86}, 095025 (2012)
  [arXiv:1207.1313 [hep-ph]].
  %%CITATION = ARXIV:1207.1313;%%
  %6 citations counted in INSPIRE as of 16 Oct 2015
%\cite{Chang:1993kw}
\bibitem{Chang:1993kw} 
  D.~Chang, W.~S.~Hou and W.~Y.~Keung,
``Two loop contributions of flavor changing neutral Higgs bosons to $\mu \rightarrow e \gamma$,''
  Phys.\ Rev.\ D {\bf 48}, 217 (1993)
  [hep-ph/9302267].
  %%CITATION = HEP-PH/9302267;%%
  %123 citations counted in INSPIRE as of 31 Aug 2015
%\cite{Baldini:2013ke}
\bibitem{Baldini:2013ke} 
  A.~M.~Baldini {\it et al.},
``MEG Upgrade Proposal,''
  arXiv:1301.7225 [physics.ins-det].
  %%CITATION = ARXIV:1301.7225;%%
  %104 citations counted in INSPIRE as of 25 Sep 2015  
%\cite{Bennett:2008dy}


%%%%%%%%%%%%%%%%%%%%%%%%%%%


\bibitem{mu-e1} 
  W.~H.~Bertl {\it et al.}  [SINDRUM II Collaboration],
  ``A Search for muon to electron conversion in muonic gold,''
  Eur.\ Phys.\ J.\ C {\bf 47}, 337 (2006).
  %%CITATION = EPHJA,C47,337;%%
  %120 citations counted in INSPIRE as of 23 May 2014


\bibitem{mu-e2} 
  H.~Natori [DeeMe Collaboration],
  ``DeeMe experiment - An experimental search for a mu-e conversion 
  reaction at J-PARC MLF,''
  Nucl.\ Phys.\ Proc.\ Suppl.\  {\bf 248-250}, 52 (2014).
  %%CITATION = NUPHZ,248-250,52;%%
  

\bibitem{mu-e3}
  Y.~Kuno [COMET Collaboration],
  ``A search for muon-to-electron conversion at J-PARC: 
  The COMET experiment,''
  PTEP {\bf 2013} (2013) 022C01.


\bibitem{mu-e4} 
  L.~Bartoszek {\it et al.}  [Mu2e Collaboration],
  %``Mu2e Technical Design Report,''
  arXiv:1501.05241 [physics.ins-det].
  %%CITATION = ARXIV:1501.05241;%%
%%%%%%%%%%%%%%%%%%%%%%%%%%%%



\bibitem{Bennett:2008dy} 
  G.~W.~Bennett {\it et al.} [Muon (g-2) Collaboration],
  ``An Improved Limit on the Muon Electric Dipole Moment,''
  Phys.\ Rev.\ D {\bf 80}, 052008 (2009)
  [arXiv:0811.1207 [hep-ex]].
  %%CITATION = ARXIV:0811.1207;%%
  %92 citations counted in INSPIRE as of 04 ao\UTF{00FB}t 2015
  %\cite{Semertzidis:1999kv}
\bibitem{Semertzidis:1999kv} 
  Y.~K.~Semertzidis {\it et al.},
  ``Sensitive search for a permanent muon electric dipole moment,''
  AIP Conf.\ Proc.\  {\bf 564}, 263 (2001)
  [hep-ph/0012087];
  %%CITATION = HEP-PH/0012087;%%
  %72 citations counted in INSPIRE as of 04 ao\UTF{00FB}t 2015
  %
  R.M.~Carey {\it et al.} ``AGS Letter of Intent: Search for a Permanent Muon Electric
  Dipole Moment'', Feb. 2000;
  A.~Silenko {\it et al.} [Muon EDM Collaboration],
  ``J-PARC Letter of Intent: Search for a Permanent Muon Electric Moment at the
  $10^{-24}~e\cdot {\rm cm}$ Level'', Jan. 2003.
%\cite{Aubert:2009qj}
\bibitem{Aubert:2009qj} 
  B.~Aubert {\it et al.} [BaBar Collaboration],
  ``Measurements of Charged Current Lepton Universality and $|V_{us}|$ using Tau Lepton Decays to $e^- \bar{\nu}_e \nu_\tau$, $\mu^-\bar{\nu}_\mu\nu_\tau$, $\pi^- \nu_\tau$ and
  $K^- \nu_\tau$,''
  Phys.\ Rev.\ Lett.\  {\bf 105}, 051602 (2010)
  [arXiv:0912.0242 [hep-ex]].
  %%CITATION = ARXIV:0912.0242;%%
  %39 citations counted in INSPIRE as of 11 Aug 2015
  %\cite{Crivellin:2013wna}
\bibitem{Crivellin:2013wna} 
  A.~Crivellin, A.~Kokulu and C.~Greub,
``Flavor-phenomenology of two-Higgs-doublet models with generic Yukawa structure,''
  Phys.\ Rev.\ D {\bf 87}, no. 9, 094031 (2013)
  [arXiv:1303.5877 [hep-ph]].
  %%CITATION = ARXIV:1303.5877;%%
  %64 citations counted in INSPIRE as of 11 sept. 2015
\bibitem{Mu3e}
  A. Blondel {\it et al.},
  ``Research Proposal for an Experiment to Search for the Decay $\mu \rightarrow eee$'',
  Dec. 2012.
%\cite{Sher:2002ew}
\bibitem{Sher:2002ew} 
  M.~Sher,
 ``tau $\to$ mu eta in supersymmetric models,''
  Phys.\ Rev.\ D {\bf 66}, 057301 (2002)
  [hep-ph/0207136].
  %%CITATION = HEP-PH/0207136;%%
  %96 citations counted in INSPIRE as of 16 Oct 2015
%\cite{Black:2002wh}
\bibitem{Black:2002wh} 
  D.~Black, T.~Han, H.~J.~He and M.~Sher,
  ``tau - mu flavor violation as a probe of the scale of new physics,''
  Phys.\ Rev.\ D {\bf 66}, 053002 (2002)
  [hep-ph/0206056].
  %%CITATION = HEP-PH/0206056;%%
  %95 citations counted in INSPIRE as of 16 Oct 2015  
  %\cite{Celis:2013xja}
\bibitem{Celis:2013xja} 
  A.~Celis, V.~Cirigliano and E.~Passemar,
``Lepton flavor violation in the Higgs sector and the role of hadronic $\tau$-lepton decays,''
  Phys.\ Rev.\ D {\bf 89}, 013008 (2014)
  [arXiv:1309.3564 [hep-ph]].
  %%CITATION = ARXIV:1309.3564;%%
  %20 citations counted in INSPIRE as of 11 sept. 2015

  %\bibitem{Cao:2005zk}
%  Q.~H.~Cao, D.~Nomura, K.~Tobe and C.~P.~Yuan,
%  %``Enhancement of 'CP-odd' Higgs boson production in the minimal
%  %supersymmetric standard model with explicit CP violation,''
%  arXiv:hep-ph/0508311.
%  %%CITATION = HEP-PH 0508311;%%
%For detail discussion and references, please see this reference.
%
%\bibitem{LightestHiggsMass}
%    Y.~Okada, M.~Yamaguchi and T.~Yanagida,
%    %``Upper Bound Of The Lightest Higgs Boson Mass In The Minimal 
%    %Supersymmetric Standard Model,''
%    Prog.\ Theor.\ Phys.\  {\bf 85}, 1 (1991);\\
%    %%CITATION = PTPKA,85,1;%%
%%
%    H.~E.~Haber and R.~Hempfling,
%    %``Can The Mass Of The Lightest Higgs Boson Of The Minimal 
%    %Supersymmetric Model Be Larger Than M(Z)?,''
%    Phys.\ Rev.\ Lett.\  {\bf 66}, 1815 (1991);\\
%    %%CITATION = PRLTA,66,1815;%%
%%
%    J.~R.~Ellis, G.~Ridolfi and F.~Zwirner,
%    %``Radiative Corrections To The Masses Of Supersymmetric Higgs Bosons,''
%    Phys.\ Lett.\ B {\bf 257}, 83 (1991).\\
%    %%CITATION = PHLTA,B257,83;%%
%
%    For recent progresses, see e.g.\\
%    S.~P.~Martin,
%    %``Two-loop scalar self-energies and pole masses in a general 
%    %renormalizable theory with massless gauge bosons,''
%    Phys.\ Rev.\ D {\bf 71}, 116004 (2005)
%    [arXiv:hep-ph/0502168];\\
%    %%CITATION = HEP-PH 0502168;%%
%%
%    S.~Heinemeyer, W.~Hollik and G.~Weiglein,
%    %``Electroweak precision observables in the minimal supersymmetric  
%    %standard model,''
%    arXiv:hep-ph/0412214\\
%    %%CITATION = HEP-PH 0412214;%%
%%
%    and references therein.
%
%
%\bibitem{ATLAS_TDR}
%    ATLAS Collaboration, 
%    ``ATLAS Detector and Physics Performance Technical Design Report'',
%    CERN/LHCC 99-14/15 (1999).
%
\end{thebibliography}
\end{document}